\documentclass[12pt, authoryear, review]{elsarticle}
\usepackage{scalefnt}
\usepackage{amssymb}
\usepackage{amsfonts}
\usepackage{amsmath}
\usepackage[nohead]{geometry}
\usepackage[singlespacing]{setspace}
\usepackage[bottom]{footmisc}
\usepackage{indentfirst}
\usepackage{endnotes}
\usepackage{graphicx}
\usepackage{rotating}

\newtheorem{theorem}{Theorem}[section]

\newtheorem{corollary}[theorem]{Corollary}

\newtheorem{lemma}[theorem]{Lemma}

\newtheorem{proposition}[theorem]{Proposition}

\newenvironment{proof}[1][Proof]{\noindent\textbf{#1.} }{\ \rule{0.5em}{0.5em}}
\makeatletter
\def\@biblabel#1{\hspace*{-\labelsep}}
\makeatother
\input{tcilatex}
\begin{document}

\begin{frontmatter}



%% \title{Decompositions of two player games: \\
%% potential, zero-sum, and stable games\tnoteref{t1}}
%% \tnotetext[t1]{This version: \today; %We thank ...
%% }

\title{Decompositions of two player games: \\
 potential, zero-sum, and stable games}

\author[A1]{Sung-Ha Hwang\corref{cor1}}
\ead{hwang@math.umass.edu, Tel: 413-545-2762, Fax:413-545-1801 }
\author[A2]{Luc Rey-Bellet}
\ead{luc@math.umass.edu }

\cortext[cor1]{Corresponding author. The research of S.-H. H. was supported by the National Science Foundation through the grant NSF-DMS-0715125.}
\address[A1]{Department of Mathematics and Statistics, University of Massachusetts Amherst, \\
Lederle Graduate Research Tower, MA 01003-9305, U.S.A.}
\address[A2]{Department of Mathematics and Statistics, University of Massachusetts Amherst, \\
Lederle Graduate Research Tower, MA 01003-9305, U.S.A.}

%\author[SHH]{Sung-Ha Hwang\corref{cor1}}
%\ead{hwang@math.umass.edu}
%\address[SHH]{Department of Mathematics and Statistics, University of Massachusetts at
%Amherst, MA, 01003-9305, U.S.A.}

\begin{abstract}

We introduce several methods of decomposition for two player normal form games. 
Viewing the set of all games as a vector space, we exhibit explicit orthonormal bases 
for the subspaces of potential games,  zero-sum games, and their orthogonal complements 
which we call anti-potential games  and  anti-zero-sum games, respectively.  
Perhaps surprisingly, every anti-potential game comes either from the \emph{Rock-Paper-Scissors} 
type games (in the case of symmetric games) or from the \emph{Matching Pennies} type games 
(in the case of asymmetric games).  Using these decompositions, we prove old (and some new) 
cycle criteria for potential and zero-sum games (as orthogonality relations between subspaces).  
We illustrate the usefulness of our decomposition by (a) analyzing the generalized 
Rock-Paper-Scissors game, (b) completely characterizing the set of all null-stable games, 
(c) providing a large class of strict stable games, (d) relating the game decomposition 
to the decomposition of vector fields for the replicator equations,  (e) constructing 
Lyapunov functions for some replicator dynamics, and (f) constructing Zeeman 
games -games with an interior asymptotically stable Nash equilibrium and a pure strategy ESS. 
\end{abstract}


% \strut\textbf{JEL Classification Numbers:} C70, D72, D74
\end{frontmatter}

\thispagestyle{empty}\onehalfspacing

\newpage

\section{Introduction\setcounter{page}{1}}

Two player normal form games (or bi-matrix games) are among the most simple
and popular games. The symmetric games in which the two players do not
distinguish between the different roles of the play have been widely used in
evolutionary dynamics, and such dynamics have been extensively studied %
\citep{Weibull95, Hofbauer98, Sandholm08}. Special classes of games such as
potential games, zero-sum games, and stable games have received a great deal
of attention because of their respective analytical advantages. For
instance, in potential games, all players' motivations to choose and deviate
from a certain strategy are described by a \emph{single} function, called a
potential function \citep{Monderer96}.

The conditions under which a game belongs to these classes have been
examined by several researchers 
\citep{Hofbauer85, Monderer96, Ui00,
Hofbauer09, Sandholm10}. For example \citet{Monderer96} and %
\citet{Hofbauer98} provide four-cycle criteria for potential games and
zero-sum games 
\citep[See Theorem 11.2.2 and Exercise 11.2.9
in][]{Hofbauer98}. Unlike existing approaches, our focus here is to examine
the extent to which a given game fails to be a potential game or a zero-sum
game.

Our basic insight is to view the set of all games as a vector space endowed
with its scalar product. Natural classes of games form subspaces of this
vector space and we systematically analyze these subspaces and their
orthogonal complements. At the very basic level it provides an immediate
intuition about the games and their dynamics. A game which consists of a
potential game plus a small non-potential part is expected, generically, by
stability to exhibit a dynamic close to the gradient-like dynamic of a
potential game. On the contrary a game with a large non-potential part will
be rather close to a volume-preserving dynamics with cycling behavior. In
addition our decomposition will clarify the relationship between potential
and zero-sum games by analyzing completely the class of games which are both
potential and zero-sum.

\begin{figure}[tb]
\centering 
\includegraphics[scale=0.4]{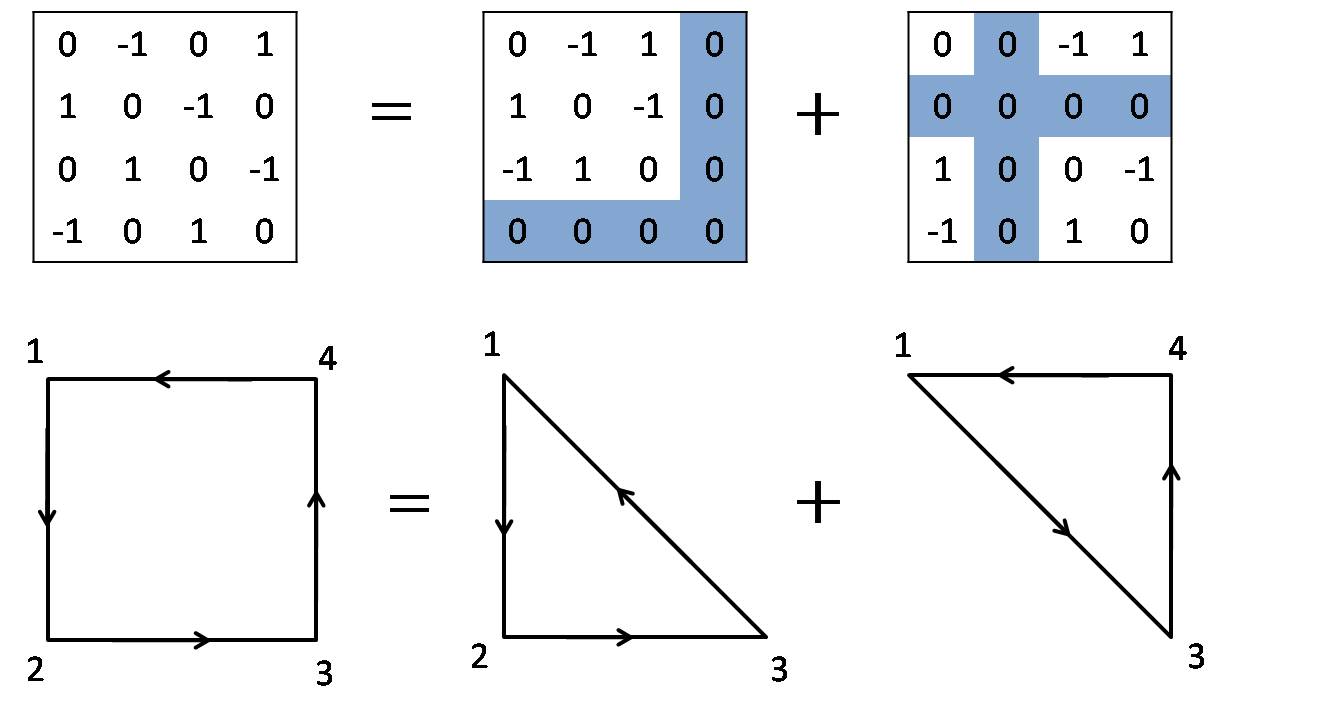}
\caption{\textbf{Decomposition of Games and Representations.} 
{\protect\footnotesize {The upper panel shows the decomposition of a
symmetric game into the two extended Rock-paper-scissors games. In the first
extended Rock-paper-scissors game, strategy 4 is \textquotedblleft
null\textquotedblright , while in the second one strategy 2 is null. The
lower panel shows the representation of these games. }}}
\label{fig:decompose}
\end{figure}

We develop three decomposition methods of bi-matrix games. In the first
decomposition, we consider the subspace of potential games and its
orthogonal complement which we call \textquotedblleft
anti-potential\textquotedblright\ games (see Figure \ref%
{fig:intro_decomposes}). Maybe surprisingly, anti-potential games are
entirely described in terms of either the \emph{Rock-Paper-Scissors} games
in the case of symmetric games, and the \emph{Matching Pennies } games in
the case of bi-matrix games. In the space of symmetric games with three
strategies, the only anti-potential game is the Rock-Paper-Scissors games,
up to a constant multiple. For symmetric games with more than three
strategies, the extended Rock-Paper-Scissors game, which involves three
strategies as Rock, Paper, and Scissors, forms a basis for the
anti-potential games (see the upper panels of Figure \ref{fig:decompose}).
Similarly, (extended) Matching Pennies games provide a basis for bi-matrix
anti-potential games.

\begin{figure}[tb]
\centering\includegraphics[scale=0.3]{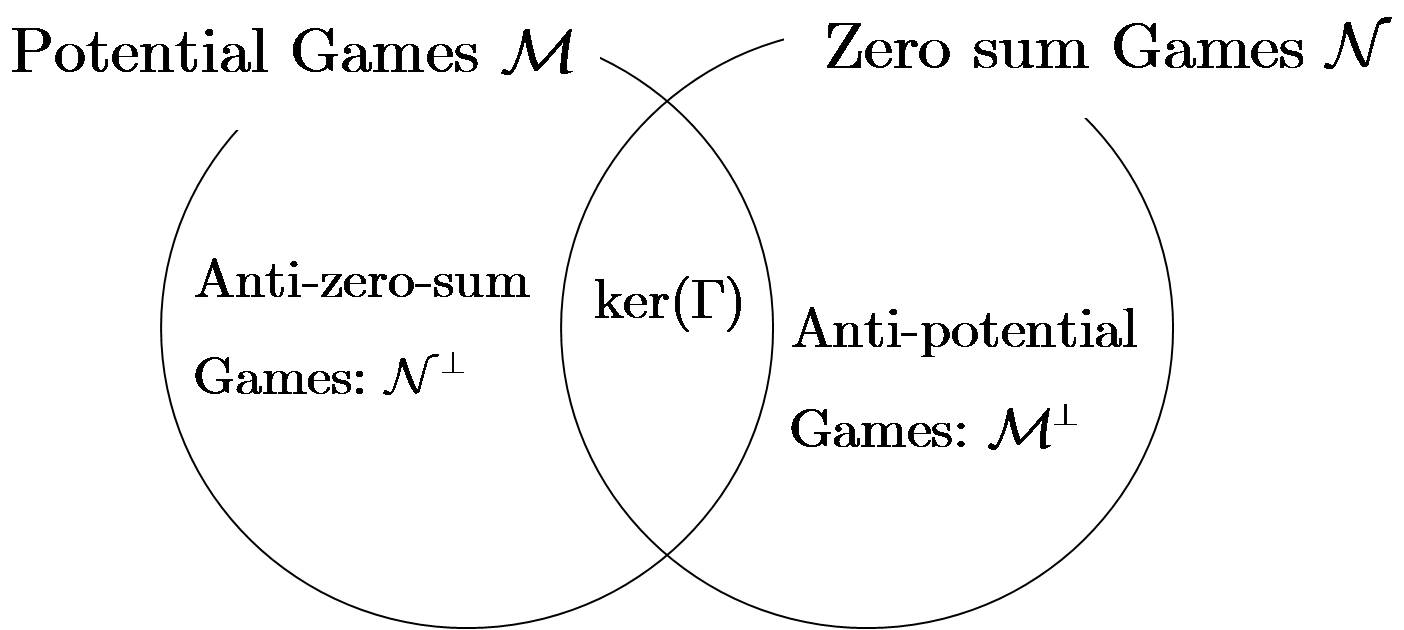}
\caption{\textbf{Decomposition Diagram.} }
\label{fig:intro_decomposes}
\end{figure}

In our second decomposition we start with the subspace of zero-sum games
(see Figure \ref{fig:intro_decomposes} again) and find that the orthogonal
complement is a special subspace of the potential game subspace (potential
games for which the sums of rows are all zero). This class of potential
games plays an important role in understanding the structures of stable
games.

Finally to understand the relationship between these two decompositions and
hence potential games and zero-sum games we use the projection mapping $P$
onto the tangent space of the simplex \citep[See][]{Sandholm08, Hofbauer09}.
We derive a third decomposition of the space of all games by considering the
mapping $\Gamma (A)=PAP$ for any matrix $A$. The kernel of this mapping turn
out to consist of games which are both zero-sum and potential games (See
Figure \ref{fig:intro_decomposes}$)$. We also show that the kernel coincides
with the set of all games with dominant strategies. Thus, the simplest
dynamics, namely the one induced by the Prisoner's Dilemma, is the only
possible type of the dynamics that games belonging to both potential and
zero-sum spaces can exhibit. In addition, we show that the subspace of all
potential games has an orthogonal decomposition into the subspace of all
anti-zero-sum games and the kernel of the mapping $\Gamma $. Similarly, the
subspace of all zero-sum games is the direct sum of the subspace of
anti-potential games and the kernel of $\Gamma $ $\ $(See Figure \ref%
{fig:intro_decomposes}$)$. This implies that the space of two player games
can be uniquely decomposed into three orthogonal subspaces:\ the subspaces
of anti-potential games, of anti-zero-sum games, and of the kernel of $%
\Gamma $. The map $\Gamma $ has been used in the recent work %
\citet{Sandholm10} on the decomposition of normal form games: our
decomposition implies that the range of $\Gamma$ can be further decomposed
into two nice classes of games; the anti-potential games and anti-zero-sum
games. %This allows us to understand the relationship between
%potential games and zero-sum games, characterize the all null-stable games,
%and identify Lyapunov functions in the dynamics induced by the underlying
%game.

We illustrate the effectiveness of these decompositions by considering
several applications:

\medskip \noindent \textbf{Algorithmic methods of identifying potential and
zero-sum games:} The decompositions provide algorithmic methods to test
whether a game is a potential game, a zero-sum game, or both. Providing
explicit bases for subspaces of games allows an easy numerical
implementation even for a large number of strategies. %
%As payoff structures of
%basis games are simple, consisting either zero or one, one can easily write
%a computer program which can decompose matrix games even for the large
%number of strategies.

\smallskip \noindent \textbf{Representation of games:} The decompositions
allow to understand easily the structure of well known games such as the
generalized Rock-Paper-Scissors game, universally cycling games %
\citep[][p.98]{Hofbauer98}. In particular, the simple anti-potential games
have a graphical representation using the orthonormal basis of games (See
the lower panels of Figure \ref{fig:decompose}).

\smallskip \noindent \textbf{Stable games:} We provide a complete
characterization of the null-stable games; every null stable game is a
zero-sum game. The converse of the statement is obvious from the definition;
using the decomposition we show that there is no null stable game which is
not a zero-sum game, a non-trivial claim. Since there are games that are
both potential and zero-sum, this shows that some potential games are
null-stable. In addition, because every bi-matrix stable game is null stable %
\citep{Hofbauer09}, this provides a complete characterization of two-player
asymmetric stable games. We also present an explicit class of strict stable
games using the decompositions.

\smallskip \noindent \textbf{Dynamics:} The decompositions are useful to
analyze the evolutionary dynamics of the underlying game. (a) Lyapunov
functions arise naturally from the perspective of decompositions. (b) Our
decomposition yields the Hodge decomposition for the vector field of the
replicator dynamics, i.e. the decomposition into a gradient-like part
(anti-zero-sum), a monotonic part (kernel of \thinspace $\Gamma $), and a
circulation part (anti-potential) (See \citet{Abraham88} and equation (1) in %
\citet{Tong03}). (c) We construct games with special properties: for example
we will explain how to construct Zeeman games, namely games with an interior
attracting fixed point and a strict pure NE (hence an ESS) and we provide
such an example for four strategy games.

These decompositions turn out to be extremely useful also for the stochastic
updating mechanism of games with finite populations. These applications will
be studied elsewhere.

%
%
%
%Furthermore wc
%\item \textbf{Hodge decomposition:} We obtain the decompositions of a given
%vector field; a gradient-like part (anti-zero-sum), a monotonic part (kernel
%of \thinspace $\Gamma $), and a circulation part (anti-potential). This
%coincides with the known decomposition of vector fields, Hodge decomposition
%(Ref here), and our decomposition reveals the systemic and fundamental
%relationship between the decomposition of vector fields and the
%decompositions of underlying games.
%\end{itemize}

This paper is organized as follows: in Section 2 we present the three
decompositions of the bi-matrix games and of the symmetric games, give
several examples and discuss also the decomposition of $n-$player normal
form games. In Section 3 we characterize stable games using the
decompositions, explain the implications of the decompositions on the
dynamics, and provide a four-strategy Zeeman game. We provide in the main
text some proofs that are important in the expositions of the paper; tedious
and book-keeping proofs are relegated to the Appendix.

\section{Decompositions of the Space of Games into Orthogonal Subspaces}

%\subsection{Reduction of Games Modulo Payoff Transformation.}

\subsection{Potential games and zero-sum games decompositions}

To illustrate the idea of our first decomposition, we decompose the
well-known generalized Rock-Paper-Scissors game by performing a simple
calculation.%
\begin{eqnarray}
&&%
\begin{pmatrix}
\gamma _{1} & -a+\gamma _{2} & b+\gamma _{3} \\ 
b+\gamma _{1} & 0+\gamma _{2} & -a+\gamma _{3} \\ 
-a+\gamma _{1} & b+\gamma _{2} & \gamma _{3}%
\end{pmatrix}
\label{eq:ex1} \\
&=&\underset{\text{Passive Game}}{\underbrace{%
\begin{pmatrix}
\gamma _{1} & \gamma _{2} & \gamma _{3} \\ 
\gamma _{1} & \gamma _{2} & \gamma _{3} \\ 
\gamma _{1} & \gamma _{2} & \gamma _{3}%
\end{pmatrix}%
}}+\frac{1}{2}(b-a)\underset{\text{Potential Part}}{\underbrace{%
\begin{pmatrix}
0 & 1 & 1 \\ 
1 & 0 & 1 \\ 
1 & 1 & 0%
\end{pmatrix}%
}}+\frac{1}{2}(b+a)\underset{\text{Anti-potential Part}}{\underbrace{%
\begin{pmatrix}
0 & -1 & 1 \\ 
1 & 0 & -1 \\ 
-1 & 1 & 0%
\end{pmatrix}%
}}  \label{eq:ex2}
\end{eqnarray}%
It is easy to see that the game (\ref{eq:ex1}) is a potential game$\ $if and
only if $a=-b$ and is equivalent to the Rock-paper-scissors game if and only
if $a=b.$ In this section we show that such a decomposition as in (\ref%
{eq:ex2}) holds for any game.

We start with symmetric games: let us denote the space of all $l\times l$
matrices by $\mathcal{L}$, and let us endow $\mathcal{L}$ with the inner
product $\left\langle A,B\right\rangle _{\mathcal{L}}=\mathrm{tr}(A^{T}B)$.
A \emph{passive game} (in the terminology of \citet{Sandholm08}) is a game
in which players' payoffs do not depend on the choice of strategies. Let $%
E_{\gamma }^{(j)}\in \mathcal{L}$ be the matrix given by 
\begin{equation*}
E_{\gamma }^{(j)}(k,l)=\left\{ 
\begin{tabular}{ll}
1 & if $k=j$ \\ 
0 & otherwise%
\end{tabular}%
\right. \,;
\end{equation*}%
i.e., $E_{\gamma }^{(j)}$\ is a matrix which has $1$'s in its $j$th column
and $0$'s at all other entries. Then the set of all symmetric passive games
is given by $\mathcal{I}:=\mathrm{span}\{E_{\gamma }^{(i)}\}_{j.}$ It is
well-known that the set of Nash equilibria for a symmetric game is left
invariant under the addition of a passive game to the payoff matrix.

%\subsection{Potential games and zero-sum games decompositions}

To characterize the spaces of all potential games and all zero-sum games, we
define the following special matrices: 
\begin{equation*}
\text{\ }K^{(ij)}=%
\begin{tabular}{llll}
& $i$-th &  & $j$-th \\ \cline{2-4}
$i$-th $\rightarrow $ & \multicolumn{1}{|c}{-1} & \multicolumn{1}{c}{$\cdots 
$} & \multicolumn{1}{c|}{1} \\ 
& \multicolumn{1}{|c}{$\vdots $} & \multicolumn{1}{c}{} & 
\multicolumn{1}{c|}{$\vdots $} \\ 
$j$-th $\rightarrow $ & \multicolumn{1}{|c}{1} & \multicolumn{1}{c}{$\cdots $%
} & \multicolumn{1}{c|}{-1} \\ \cline{2-4}
\end{tabular}%
,\text{ }N^{(ij)}=%
\begin{tabular}{llllll}
& 1st &  & $i$-th &  & $j$-th \\ \cline{2-6}
1st & \multicolumn{1}{|c}{0} & \multicolumn{1}{c}{$\cdots $} & 
\multicolumn{1}{c}{-1} & \multicolumn{1}{c}{$\cdots $} & \multicolumn{1}{c|}{
1} \\ 
& \multicolumn{1}{|c}{$\vdots $} & \multicolumn{1}{c}{} & \multicolumn{1}{c}{%
$\vdots $} & \multicolumn{1}{c}{} & \multicolumn{1}{c|}{$\vdots $} \\ 
$i$-th $\rightarrow $ & \multicolumn{1}{|c}{1} & \multicolumn{1}{c}{$\cdots $%
} & \multicolumn{1}{c}{0} & \multicolumn{1}{c}{$\cdots $} & 
\multicolumn{1}{c|}{-1} \\ 
& \multicolumn{1}{|c}{$\vdots $} & \multicolumn{1}{c}{} & \multicolumn{1}{c}{%
$\vdots $} & \multicolumn{1}{c}{} & \multicolumn{1}{c|}{$\vdots $} \\ 
$j$-th $\rightarrow $ & \multicolumn{1}{|c}{-1} & \multicolumn{1}{c}{$\cdots 
$} & \multicolumn{1}{c}{1} & \multicolumn{1}{c}{$\cdots $} & 
\multicolumn{1}{c|}{0} \\ \cline{2-6}
\end{tabular}%
\text{ ,}
\end{equation*}%
where all other elements in the matrices are zeros. Note that $N^{(ij)}$ is
a game whose restriction on the strategy set $\{1,i,j\}\times \{1,i,j\}$ is
the Rock-Paper-Scissors game. 
%TCIMACRO{\TeXButton{\citet{Monderer96}}{\citet{Monderer96}}}%
%BeginExpansion
\citet{Monderer96}%
%EndExpansion

Recall that a symmetric game $A$ is a \emph{potential game} (%
\citet{Monderer96}) if there exist a symmetric matrix $S$ and a passive game 
$\sum_{j}\gamma _{j}E_{\gamma }^{(j)}\in \mathcal{I}$ such that 
\begin{equation}
A=S+\text{ }\sum_{j}\gamma _{j}E_{\gamma }^{(j)}\,.  \label{eq:pot}
\end{equation}%
We will use the word \textquotedblleft exact \textquotedblright\ to indicate
that a game is a potential game with no passive part, i.e., all $\gamma
_{j}=0$ (exact potential games are called full potential games in %
\citet{Sandholm08}). We denote by $\mathcal{M}$ the linear subspace of all
potential games and we have the orthogonal decomposition $\mathcal{L}=%
\mathcal{M}\oplus \mathcal{M}^{\bot }$ with respect to the inner product $%
<,>_{\mathcal{L}}$. We call a game in $\mathcal{M}^{\bot }$ an \emph{%
anti-potential game}.

Note that the dimension of the subspace of $\mathcal{L}$ consisting of all
symmetric matrices is $\frac{1}{2}l(l+1)$ and the dimension of the subspace
of passive games is $l$. Since the sum of all $E_{\gamma }$ is an exact
potential game, namely the game whose payoffs are all 1's, the dimension of
the intersection between the subspace of all symmetric matrices and $%
\mathcal{I}$ is at least $1$. Conversely if a matrix belongs to this
intersection, then the entries of this matrices should be all the same (see
also the discussion in \citet[][p.15]{Sandholm10}) and so the dimension of
the intersection is exactly $1$. Hence the dimension of $\mathcal{M}$ is
given by%
\begin{equation}
\dim (\mathcal{M)}=\frac{l(l+1)}{2}+l-1=l^{2}-\frac{(l-1)(l-2)}{2}.
\label{eq:dim}
\end{equation}%
Note that the extended Rock-Paper-Scissors game, $N^{(ij)}$, is an
anti-symmetric matrix whose column sums and row sums are all $0$'s. Thus, we
have 
\begin{equation*}
\left\langle A,N^{(ij)}\right\rangle _{\mathcal{L}}=0\text{ }\,,
\end{equation*}%
for all $A\in \mathcal{M},$ because $\left\langle S,N^{(ij)}\right\rangle _{%
\mathcal{L}}$ $=0$ and $\left\langle P,N^{(ij)}\right\rangle _{\mathcal{L}%
}=0 $ for all symmetric matrix $S$ and all passive game $P$ (See the
appendix for the properties of $\left\langle {}\right\rangle _{\mathcal{L}%
}). $ \ In other words, $N^{(ij)}\in \mathcal{M}^{\perp }$ for all $i,j$.
The set $\{N^{(ij)}:j>i,\text{ }i=2,\cdots ,l-1\}$ has $\frac{(l-1)(l-2)}{2}$
elements and they are linearly independent since each $N^{(ij)}$ is uniquely
determined by the property of having $1$ in its $(i,j)$ th position. This
set forms a basis for $\mathcal{M}^{\perp }.$ If \ a matrix $B $ is
antisymmetric and the sums of elements in each column in $B$ are all zeros, $%
\ \left\langle S,B\right\rangle _{\mathcal{L}}=0$ for a symmetric matrix and 
$\left\langle P,B\right\rangle _{\mathcal{L}}=0$ for a passive game $P.$ \
Therefore $B\in $ $\mathcal{M}^{\perp }$. On the other hand, $\ $if $B\in 
\mathcal{M}^{\perp }$, $\ B$ can be written as a linear combination of $%
N^{(ij)}$, and hence $B$ is antisymmetric and the sums of elements in each
column in $B$ are all zeros. Thus we obtain

\begin{proposition}[Anti-potential games]
\label{prop_sym_m_per} We have 
\begin{equation*}
B\in \mathcal{M}^{\perp }\text{ if and only if \ }B^{T}=-B\mathrm{~and~}%
\sum_{j}B(i,j)=\sum_{i}B(i,j)=0\,.
\end{equation*}%
Moreover the set $\{N^{(ij)}:j>i,$ $i=2,\cdots ,l\}$ forms a basis for $%
\mathcal{M}^{\perp }$.
\end{proposition}

Proposition \ref{prop_sym_m_per} shows that a basis for $\mathcal{M}^{\perp
} $ can be obtained from the extended Rock-Paper-Scissors. As a corollary of
Proposition \ref{prop_sym_m_per} we obtain immediately the criterion for
potential games given by \citet{Hofbauer98}.

\begin{corollary}[Potential games]
\label{cor:pot} $A$ is a potential game if and only if 
\begin{equation}
a(l,m)-a(k,m)+a(k,l)-a(m,l)+a(m,k)-a(l,k)=0\text{ \ for all }l,m,k\in S
\label{eq:pot-sym_cri1}
\end{equation}
\end{corollary}

\begin{proof}
First note from Proposition \ref{prop_sym_m_per} that $A$ is a potential
game if and only if $\left\langle A,N^{(ij)}\right\rangle _{\mathcal{L}}=0$
for all $i,j.$ Then note that 
\begin{equation}
a(l,m)-a(k,m)+a(k,l)-a(m,l)+a(m,k)-a(l,k)=\left\langle A,E\right\rangle _{%
\mathcal{L}}  \notag
\end{equation}%
where%
\begin{equation*}
E=%
\begin{tabular}{cccc}
& $k$ & $l$ & $m$ \\ \cline{2-4}
$k$ & \multicolumn{1}{|c}{0} & 1 & \multicolumn{1}{c|}{-1} \\ 
$l$ & \multicolumn{1}{|c}{-1} & 0 & \multicolumn{1}{c|}{1} \\ 
$m$ & \multicolumn{1}{|c}{1} & -1 & \multicolumn{1}{c|}{0} \\ \cline{2-4}
\end{tabular}%
\text{ and all other entries in }E\text{ are }0\text{'s.}
\end{equation*}%
Then clearly (\ref{eq:pot-sym_cri1}) implies $\left\langle
A,N^{(ij)}\right\rangle _{\mathcal{L}}=0$ for all $i,j.$ Conversely, the
matrix $E$ is anti-symmetric and its row sums and column sums are zero, so $%
E\in \mathcal{M}^{\perp }.$ Therefore $E$ can be uniquely written as $%
N^{(ij)}$ and thus $\left\langle A,N^{(ij)}\right\rangle _{\mathcal{L}}=0$
for all $i,j$ implies (\ref{eq:pot-sym_cri1}).
\end{proof}

We provide next a similar decomposition starting with zero-sum games. We
call an anti-symmetric matrix $A$ an \emph{exact zero-sum game} and call a
game \emph{zero-sum} if it can be written as the sum of a antisymmetric
matrix and a passive game. Let us denote by $\mathcal{N}$ the subspace of
all zero-sum games. The dimension of the subspace all anti-symmetric
matrices is $\frac{(l-1)l}{2}$ %\begin{equation*}
%\text{dimension of all symmetric matrices = }\frac{(l-1)l}{2}.
%\end{equation*}%
and the dimension of the intersection between the subspace of anti-symmetric
matrices and $\mathcal{I}$ is $0$ (the diagonal elements of anti-symmetric
matrices are all zeros and hence all off-diagonal elements are again all
zeros if this game is also a passive game). Thus 
\begin{equation}
\dim (\mathcal{N})=\frac{(l-1)l}{2}+l=l^{2}-\frac{(l-1)l}{2}.
\label{eq:dim2}
\end{equation}%
We decompose the space of game as ${\mathcal{L}}={\mathcal{N}}\oplus 
\mathcal{N}^{\perp }$ and we call a game in $\mathcal{N}^{\perp }$ an \emph{%
anti-zero-sum} game. Note that $K^{(ij)}$ is a symmetric matrix whose row
sums and column sums are zeros, so $K^{(ij)}\in \mathcal{N}^{\perp }.$ The
set $\{K^{(ij)}:j>i,\text{ }i=1,\cdots ,l\}$ has $\frac{(l-1)l}{2}$ elements
which are linearly independent since each $K^{(ij)}$ is uniquely determined
by having $1$ in its $(i,j)$th entry. Thus we obtain

%
%We consider
%the following set of $K^{(ij)}$'s :%
%\begin{equation}
%\{K^{(ij)}:j>i,\text{ }i=1,\cdots ,l\}.  \label{eq:basis_N}
%\end{equation}%
%Then the cardinality of (\ref{eq:basis_N}) is $\frac{(l-1)l}{2}$ and all
%elements in ($\ref{eq:basis_N})$ are linearly independent since each $%
%K^{(ij)}$ is uniquely determined by having $1$ at its $(i,j)$th position.

\begin{proposition}[Anti-zero-sum games]
\label{prop:sym_char_n} We have 
\begin{equation*}
B\in \mathcal{N}^{\perp }\text{ }\ \text{if and only if \ }B^{T}=B\mathrm{%
~and~}\sum_{j}B(i,j)=\sum_{i}B(i,j)=0\,.
\end{equation*}%
Moreover the set $\{K^{(ij)}:j>i,$ $i=1,\cdots ,l-1\}$ forms a basis for $%
\mathcal{N}^{\perp }$.
\end{proposition}

%Therefore we obtain another decomposition of $\mathcal{L},$ $\mathcal{L}=%
%\mathcal{N}\oplus \mathcal{N}^{\bot },$ and call a game $A\in \mathcal{N}%
%^{\perp }$ an anti-zero-sum game. 

Using this orthogonal decomposition we obtain a new criterion to identify a
zero-sum game similar to the criterion in Corollary \ref{cor:pot}.

\begin{corollary}[Zero-sum games]
$A$ is a zero-sum game if and only if 
\begin{equation}  \label{zerosumcrit}
a(j,i)-a(i,i)+a(i,j)-a(j,j)=0 \mathrm{~for~ all~} i,j\in S \,.
\end{equation}
\end{corollary}

\begin{proof}
If $A \in {\mathcal{N}}$ then $\left\langle A,K^{(ij)}\right\rangle _{%
\mathcal{L}}=0$ which yields (\ref{zerosumcrit}). %\begin{equation*}
%\left\langle A,K^{(ij)}\right\rangle _{\mathcal{L}%
%}=-a(i,i)+a(j,i)-a(j,i)+a(i,j),
%\end{equation*}
\end{proof}

\subsection{Decomposition using the projection mapping $\Gamma$}

The subspaces of potential games and zero-sum games have a non-trivial
intersection ${\mathcal{M}}\cap {\mathcal{N}}$. In order to understand this
set let $P=I-\frac{1}{l}\mathbf{11}^{T}$ where $I$ is the identity matrix
and $\mathbf{1}$ the constant vector with entries equal to 1. It is easy to
see that $P$ is the orthogonal projection onto the subspace $T\Delta= \{x\in 
\mathbb{R}^{l}\,;\sum_{i}x_{i}=0\}$, i.e., onto the tangent space to the
unit simplex $\Delta= \{x\in \mathbb{R}^{l}\,;x_{i}\geq
0\,,\sum_{i}x_{i}=1\} $. Let us define a linear transformation $\Gamma $ on $%
\mathcal{L}$ by 
\begin{equation*}
\Gamma :\mathcal{L\rightarrow L},\text{ }A\mapsto PAP.
\end{equation*}

%We can understand the previous two decompositions as decompositions of the
%space $\mathcal{L}$ into the kernel of the orthogonal projection $($in $%
%\mathcal{L})$ onto $\mathcal{M}$ or $\mathcal{N}$ and the range of this
%projection map. Similarly, we\ define a linear transformation $\Gamma $ on $%
%\mathcal{L}$:%
%\begin{equation*}
%\Gamma :\mathcal{L\rightarrow L},\text{ }A\mapsto PAP,\text{ }P=I-\frac{1}{l}%
%\mathbf{11}^{T}.
%\end{equation*}%
%Here the matrix $P$ is the usual orthogonal projection matrix onto the
%tangent space of the simplex. 

To characterize the kernel and the range of the map $\Gamma $, let us say
that a game is a \emph{constant game} if the player's payoff does not depend
on his opponent's strategy, that is the payoff matrix is constant on each
row. The matrices $E_{\eta }^{(i)}:=(E_{\gamma }^{(i)})^{T}$ form an
orthonormal basis of the subspace of constant games. Note that $E_{\eta
}^{(i)}$ has a strictly dominant strategy. Furthermore let us define 
%for each $i\in \{2,\cdots ,l_{r}\},$\textsl{\ }$%
%j\in \{2,\cdots ,l_{c}\}$%
\begin{equation*}
E_{\kappa }^{(ij)}=%
\begin{tabular}{ccccc}
&  & $j-$th & $j+1-$th &  \\ \cline{2-5}
& \multicolumn{1}{|c}{} & $\vdots $ & $\vdots $ & \multicolumn{1}{c|}{} \\ 
$i-$th & \multicolumn{1}{|c}{$\cdots $} & -1 & 1 & \multicolumn{1}{c|}{$%
\cdots $} \\ 
$i+1-$th & \multicolumn{1}{|c}{$\cdots $} & 1 & -1 & \multicolumn{1}{c|}{$%
\cdots $} \\ 
& \multicolumn{1}{|c}{} & $\vdots $ & $\vdots $ & \multicolumn{1}{c|}{} \\ 
\cline{2-5}
\end{tabular}%
\text{ where all other entries are }0\text{'s.}
\end{equation*}
It is easy to see that 
\begin{equation}
\mathrm{span}\{E_{\eta }^{(1)},\cdots ,E_{\eta }^{(l)},E_{\gamma
}^{(1)},\cdots ,E_{\gamma }^{(l)}\}\subset \ker \Gamma .  \label{eq:span}
\end{equation}%
Conversely, one can show that the left and right actions of the projection
matrices makes only this class belongs to $\ker \Gamma $. $\ $Then note that%
\begin{equation*}
\sum_{i}E_{\gamma }^{(l)}=\sum_{i}E_{\eta }^{(l)},
\end{equation*}%
so by throwing away one element from the spanning set (\ref{eq:span}), we
may obtain the independent spanning set, hence a basis for the kernel of $%
\Gamma .$ Concerning the range of $\Gamma ,$ by counting the basis elements,
we have dim($\ker \Gamma )=2l-1$ and, thus, dim(range$\Gamma
)=l^{2}-(2l-1)=(l-1)^{2}.$ Since $\mathbf{1}E_{\kappa }^{(ij)}=\mathbf{0}$
and $E_{\kappa }^{(ij)}\mathbf{1}=\mathbf{0},$ 
\begin{equation*}
\{E_{\kappa }^{(ij)}:i=1,\cdots ,l-1,j=1,\cdots l-1\}
\end{equation*}%
provides a natural candidate for the basis of the range. These observations
lead to Proposition $\ref{prop:char_ker}$ whose formal proof is elementary
but tedious, and hence relegated to the Appendix.

\begin{proposition}[Characterizations of $\ker (\Gamma )$ and range($\Gamma
) $]
\label{prop:char_ker} We have \newline
(1) $\{E_{\eta }^{(i)}\}_{i\neq 1}\cup \{E_{\gamma }^{(j)}\}_{j}$ form a
basis for $\ker \Gamma .$\newline
(2) $\{E_{\kappa }^{(ij)}:i=1,\cdots ,l-1,j=1,\cdots l-1\}$ form a basis for
range$(\Gamma )$.
\end{proposition}

Next, we study the relationship among these subspaces. First every game in
the subspace $\mathcal{N}^{\perp }$ is a symmetric matrix and thus a
potential game. Similarly every anti-potential game is a zero-sum game, so
we have $\mathcal{N}^{\perp }\subset $~$\mathcal{M}$ and $\mathcal{M}^{\perp
}\subset \mathcal{N}.$ To understand the relationship among these spaces
further, note the following facts:%
\begin{eqnarray*}
\begin{pmatrix}
1 & 1 & 1 \\ 
0 & 0 & 0 \\ 
0 & 0 & 0%
\end{pmatrix}%
+%
\begin{pmatrix}
1 & 0 & 0 \\ 
1 & 0 & 0 \\ 
1 & 0 & 0%
\end{pmatrix}
&=&%
\begin{pmatrix}
2 & 1 & 1 \\ 
1 & 0 & 0 \\ 
1 & 0 & 0%
\end{pmatrix}
\\
\text{ }%
\begin{pmatrix}
1 & 1 & 1 \\ 
0 & 0 & 0 \\ 
0 & 0 & 0%
\end{pmatrix}%
-%
\begin{pmatrix}
1 & 0 & 0 \\ 
1 & 0 & 0 \\ 
1 & 0 & 0%
\end{pmatrix}
&=&%
\begin{pmatrix}
0 & 1 & 1 \\ 
-1 & 0 & 0 \\ 
-1 & 0 & 0%
\end{pmatrix}%
.\text{\ }
\end{eqnarray*}%
This example illustrates the fact that any game in $\ker (\Gamma )$ which is
not a passive game is both a potential game and zero-sum game; i.e., every
constant game is both a potential games and zero-sum games. As Proposition %
\ref{prop-space} shows the converse holds: a game which is both a potential
and a zero-sum game is equivalent to a constant game.

\begin{proposition}
\label{prop-space}$\ker (\Gamma )=\mathcal{M}$ $\mathcal{\cap }$ $\mathcal{N}
$ and range$(\Gamma )=\mathcal{M}^{\perp }\oplus \mathcal{N}^{\perp }.$
\end{proposition}

Proposition \ref{prop-space} provides the essential characterization of the
relationship among spaces. Since $\mathcal{L}=\ker (\Gamma )\oplus \mathrm{%
range}(\Gamma ),$ from Proposition \ref{prop-space}, we obtain the
decomposition of a given game into three parts; $\mathcal{L=M}^{\perp
}\oplus \mathcal{N}^{\perp }\oplus \ker (\Gamma \mathbf{).}$ Also since $%
\mathcal{N}\cap (\mathcal{M}^{\perp }\mathcal{\cup N}^{\perp }\mathcal{)=M}%
^{\perp },$ we will have $\mathcal{N}\cap $ \textrm{range}$(\Gamma )=%
\mathcal{M}^{\perp }$ and this provides another characterization of $%
\mathcal{M}^{\perp }$ as follows. From Proposition \ref{prop_sym_m_per}, we
know that a game is anti-potential if and only if it is an antisymmetric
matrix whose row sums and column sums are zeros. We know that all row sums
and column sums of games belonging to range$(\Gamma )$ are zeros and the
zero sum game is the sum of an antisymmetric matrix and a passive game; thus
we can show that $\mathcal{M}^{\perp }=\mathcal{N}\cap $ \textrm{range}$(%
\mathbf{\Gamma }).$ In this way we obtain the following key result in the
paper.

\begin{theorem}
\label{cor:decomp}We have\newline
(1) $\mathcal{M}=$ $\mathcal{N}^{\perp }\oplus \ker (\Gamma )$ and $\mathcal{%
M}^{\perp }=\mathcal{N}\cap $ \textrm{\textrm{range}}$(\mathbf{\Gamma })$%
\newline
(2) $\mathcal{N}=\mathcal{M}^{\perp }\oplus \ker (\Gamma )$\ and $\mathcal{N}%
^{\perp }=\mathcal{M}\cap $ \textrm{\textrm{range}}$(\mathbf{\Gamma })$%
\newline
(3) $\mathcal{L}=\mathcal{M}^{\perp }\oplus \mathcal{N}^{\perp }\oplus \ker
(\Gamma \mathbf{)}$
\end{theorem}

\begin{proof}
(1) From Proposition \ref{prop-space}, $\ $we have $\mathcal{N}^{\perp
}+\ker (\Gamma )=\mathrm{span}(\mathcal{N}^{\perp }\cup \ker (\Gamma ))=%
\mathrm{span}((\mathcal{N}^{\perp }\cup \mathcal{M})\cap (\mathcal{N}^{\perp
}\cup \mathcal{N}))=\mathcal{M}.$ Since $\mathcal{N}^{\perp }\perp \ker
(\Gamma ),$ we have $\mathcal{M}=$ $\mathcal{N}^{\perp }\oplus \ker (\Gamma
).$ From proposition \ref{prop-space}, we have $\mathcal{M}^{\perp }\subset $
$\mathcal{M}^{\perp }\oplus \mathcal{N}^{\perp }$ = \textrm{range}$(\Gamma )$
and see that $\mathcal{M}^{\perp }\subset \mathcal{N}\cap $ \textrm{range}$%
(\Gamma ).$ Conversely again from proposition \ref{prop-space}, we have 
\begin{equation*}
\mathcal{N}\cap \text{range}(\Gamma )\newline
=\mathcal{N}\cap (\mathrm{span}(\mathcal{M}^{\perp }\cup \mathcal{N}^{\perp
}))\supset \mathrm{span}(\mathcal{N\cap }(\mathcal{M}^{\perp }\cup \mathcal{N%
}^{\perp }))=\mathcal{M}^{\perp }.
\end{equation*}%
By changing the roles of $\mathcal{M}$ and $\mathcal{N},$ we obtain (2). (3)
follows from $\mathcal{L=M}^{\perp }\oplus \mathcal{M=M}^{\perp }\oplus 
\mathcal{N}^{\perp }\oplus \ker (\Gamma ).$
\end{proof}

\citet{Sandholm10} provides a method of decomposing normal form games by
using the orthogonal projection $P:$ for a given $A$ write 
\begin{equation}
A=\underset{\in \text{ range(}\Gamma )}{\underbrace{PAP}}+\text{ }\underset{%
\in \text{ ker(}\Gamma )}{\underbrace{(I-P)AP+PA(I-P)+(I-P)A(I-P)}}.
\label{eq:sand_decomp}
\end{equation}%
The first term in (\ref{eq:sand_decomp}) belongs to the range of $\Gamma $
and the remaining three terms belong to the kernel of $\Gamma $. Our
decompositions (Proposition \ref{prop-space}) show that $PAP$ can be further
decomposed into games having nice properties $-$ potential games and
zero-sum games $-$ and every game in $\ker(\Gamma)$ is a game which is both
a potential and a zero-sum game and possesses (generically) a dominant
strategy.

Theorem \ref{cor:decomp} also provides a convenient way to compute the
anti-zero-sum part (anti-potential part, resp.) of a game when the
anti-potential part (anti-zero-sum part, resp.) is known. Suppose that $A$
is a symmetric game and its anti-potential part is $Z$. Then the part of $A$
that belongs to $\ker (\Gamma )$ is $A-PAP$. Hence from (3) of Theorem \ref%
{cor:decomp}, its anti-zero-sum part is given by $A-Z-(A-PAP)=PAP\mathbb{-}%
Z; $ in fact Theorem \ref{cor:decomp} shows that $PAP\mathbb{-}Z$ is a
symmetric matrix in $\mathcal{L}$ and its all row sums and column sums are
zeros.

\subsection{Decompositions of bi-matrix games}

In this section we prove a decomposition theorem for bi-matrix games and
elucidate the relations between the decomposition of symmetric and bi-matrix
games. Most results in this section generalize the corresponding results in
sections 2.1-2.2. We denote (with a slight abuse of notation) by $\mathcal{L}
$ the space of all $l_{r}\times l_{c}$ matrices with the inner product $%
\left\langle A,B\right\rangle _{\mathcal{L}}:=tr(A^{T}B).$ 
%Without loss of generality we
%assume $l_{r}\leq l_{c}$. 
The set of all bi-matrix games is $\mathcal{L}^{2}:=\mathcal{L\times L}$ and
sometimes we will view a bi-matrix game $(A,B)$ as a $(l_{r}+l_{c})\times
(l_{r}+l_{c})$ matrix given by 
\begin{equation*}
(A,B):=%
\begin{pmatrix}
O_{r} & A \\ 
B^{T} & O_{c}%
\end{pmatrix}%
\end{equation*}%
where $O_{r}$ and $O_{c}$ are $l_{r}\times l_{r}$ and $\ l_{c}\times l_{c}$
zero matrices, respectively. The space $\mathcal{L}^{2}$ is a linear
subspace of the set of all $(l_{r}+l_{c})\times (l_{r}+l_{c})$ matrices of
dimension $2l_{r}l_{c}$. We endow $\mathcal{L}^{2}$ with the inner product $%
<\cdot ,\cdot >_{\mathcal{L}^{2}}$, where $\left\langle
(A,B),(C,D)\right\rangle _{\mathcal{L}^{2}}:=tr((A,B)^{T}(C,D))$. The
elementary properties of this scalar product are summarized in the Appendix.

The set of all bi-matrix \emph{passive games} $\mathcal{\bar{I}}$ is given
by 
\begin{equation*}
\mathcal{\bar{I}}:=\mathrm{span}(\{(E_{\gamma }^{(j)},O)\}_{j}\cup
\{(O,E_{\gamma }^{(i)})\}_{i}).
\end{equation*}%
and we say that the games $(A,B)$ and$(C,D)$ are equivalent if $%
(A,B)-(C,D)\in \mathcal{\bar{I}}$. In this case we write $(A,B)\sim (C,D)$.
The set of Nash equilibria for a bi-matrix game is invariant under this
equivalence relation.

Note that $(E_{\kappa }^{(ij)},-E_{\kappa }^{(ij)})$ is a game whose
restriction on the strategy set $\{i,i+1\}\times \{j,j+1\}$ is the Matching
Pennies game and we call it %$(E_{\kappa }^{(ij)},-E_{\kappa }^{(ij)})$ is 
an extended Matching Pennies game.

From 
%TCIMACRO{\TeXButton{\citet{Monderer96}}{\citet{Monderer96}} }%
%BeginExpansion
\citet{Monderer96}
%EndExpansion
we recall that $(A,B)$ is a \emph{potential game} if there exist a matrix $S$
and $\{\gamma _{j}\}_{j}$ , $\{\eta _{i}\}_{i}$ such that%
\begin{equation*}
(A,B)=(S,S)+\sum_{j}\gamma _{j}(E_{\gamma }^{(j)},O)+\sum_{i}\eta
_{i}(O,E_{\eta }^{(i)})\,.
\end{equation*}%
Denoting by $\mathcal{\bar{M}}$ the subspace of all potential games, we have
the orthogonal decomposition $\mathcal{L}^{2}=\mathcal{\bar{M}}\oplus 
\mathcal{\bar{M}}^{\bot }$. The dimension of the subspace of all exact
potential games is $l_{r}\times l_{c}$ and the dimension of the subspace of
all passive games is $l_{r}+l_{c}$. Arguing as for symmetric games, one
finds that the dimension of $\mathcal{\bar{M}}$ is given by%
\begin{equation}
\dim (\mathcal{\bar{M})}%
=l_{r}l_{c}+l_{r}+l_{c}-1=2l_{r}l_{c}-(l_{r}-1)(l_{c}-1).  \label{eq:dim_bi}
\end{equation}%
Note also that $(E_{\kappa }{}^{(ij)},-E_{\kappa }{}^{(ij)})$ is an
anti-symmetric matrix as an element in $\mathcal{L}^{2}$ whose column sum
and row sum are all $0$'s, thus we have $\left\langle (A,B),(E_{\kappa
}{}^{(i,j)},-E_{\kappa }{}^{(i,j)})\right\rangle _{\mathcal{L}^{2}}=0$ for
all $(A,B)\in \mathcal{\bar{M}}.$ In other words, $(E_{\kappa
}{}^{(ij)},-E_{\kappa }{}^{(ij)})\in \mathcal{\bar{M}}^{\perp }$ for all $%
i,j $ and the number of such $(E_{\kappa }{}^{(i,j)},-E_{\kappa }{}^{(i,j)})$
is $(l_{r}-1)(l_{c}-1)$. Hence we have 
%From a counting argument  and  (\ref{eq:dim_bi}), 
%we have %$\left\{ (E_{\kappa }{}^{(i,j)},-E_{\kappa}{}^{(i,j)})\right\} _{i,j}$ can be a basis for $\mathcal{\bar{M}}^{\perp }.$

\begin{proposition}[Anti-potential games]
The set \label{prop:char_bi_m} $\ \{(E_{\kappa }{}^{(i,j)},-E_{\kappa
}{}^{(i,j)})\}_{1\le i < l_r,1\le j < l_c}$ is a basis for $\mathcal{\bar{M}}%
^{\perp }$.
\end{proposition}

\begin{proof}
From the discussion above, it is enough to show the linear independence
among $(E_{\kappa }^{(ij)},-E_{\kappa }^{(ij)}).$ To do this, we consider
the following linear combination:%
\begin{equation*}
\sum_{i,j}\kappa ^{(ij)}E_{\kappa }^{(ij)}=0.
\end{equation*}%
Then, it is easy to see that $\kappa ^{(11)}=0.$ This implies $\kappa
^{(1,j)}=0$ for all $j$ $~$which, in turn, implies $\kappa ^{(i,j)}=$ 0 for
all $i.$
\end{proof}

Proposition \ref{prop:char_bi_m} shows that a basis for $\mathcal{M}^{\perp
} $ can be obtained from the Matching Pennies games and its extensions. From
this, we say that $\left( A,B\right) $ is an bi-matrix \emph{anti-potential}
game whenever $(A,B)\in \mathcal{M}^{\perp }.$ Proposition \ref%
{prop:char_bi_m} provides an alternative and simple proof for the well-known
criterion for the potential game by \cite{Monderer96}:

\begin{corollary}[Potential games]
\label{cor:bi_pot}$(A,B)$ is a potential game if and only if for all $%
i,i^{\prime }\in S_{r}$, $j,j^{\prime }\in S_{c},$%
\begin{equation*}
a(i^{\prime },j)-a(i,j)+b(i^{\prime },j^{\prime })-b(i^{\prime
},j)+a(i,j^{\prime })-a(i^{\prime },j^{\prime })+b(i,j)-b(i,j^{\prime })=0
\end{equation*}
\end{corollary}

\begin{proof}
It is enough to notice that 
\begin{eqnarray*}
&&a(i^{\prime },j)-a(i,j)+b(i^{\prime },j^{\prime })-b(i^{\prime
},j)+a(i,j^{\prime })-a(i^{\prime },j^{\prime })+b(i,j)-b(i,j^{\prime }) \\
&=&\left\langle (A,B),(K^{(i,i^{\prime })(j,j^{\prime })},-K^{(i,i^{\prime
})(j,j^{\prime })})\right\rangle _{\mathcal{L}^{2}}
\end{eqnarray*}%
where $(K^{(i,i^{\prime })(j,j^{\prime })},-K^{(i,i^{\prime })(j,j^{\prime
})})$ is an extended Matching Pennies game whose restriction on $\left\{
i,i^{\prime }\right\} \times \{j,j^{\prime }\}$ is a Matching Pennies game.
\end{proof}

%\begin{corollary}[Potential games]
%\label{cor:bi_pot}$(A,B)$ is a potential game if and only if for for all $%
%i,i^{\prime }\in S_{r}$, $j,j^{\prime }\in S_{c},$%
%\begin{equation*}
%a(i^{\prime },j)-a(i,j)+b(i^{\prime },j^{\prime })-b(i^{\prime
%},j)+a(i,j^{\prime })-a(i^{\prime },j^{\prime })+b(i,j)-b(i,j^{\prime })=0
%\end{equation*}
%\end{corollary}

Next we consider a decomposition using zero-sum games as in Section 2.2. \
We call a game of the form $(A,-A)$ \ an exact zero-sum game and say that a
game is a \emph{zero-sum} game if it can be written as the sum of an exact
zero-sum game and a passive game. We denote by $\mathcal{\bar{N}}$ the
subspace of all bi-matrix zero-sum games and have dim($\mathcal{\bar{N}}%
)=2l_{r}l_{c}-(l_{r}-1)(l_{c}-1)$. A similar argument as in Section 2.2
yields

\begin{proposition}[Anti-zero-sum games]
The set \label{prop:char_n}$\{(E_{\kappa }^{(ij)},E_{\kappa
}^{(ij)})\}_{1\le i < l_r , 1 \le j < l_c}$ is a basis for $\mathcal{\bar{N}}%
^{\perp }.$
\end{proposition}

Again the following corollary is an immediate consequence of orthogonality %
\citep[See Exercise 11.2.9  in][]{Hofbauer98}.

\begin{corollary}[Zero-sum games]
\label{cor:bi_zero}$(A,B)$ is a zero-sum game if and only if for all $%
i,i^{\prime }\in S_{r}$, $j,j^{\prime }\in S_{c},$%
\begin{equation*}
a(i^{\prime },j)-a(i,j)-b(i^{\prime },j^{\prime })+b(i^{\prime
},j)+a(i,j^{\prime })-a(i^{\prime },j^{\prime })-b(i,j)+b(i,j^{\prime })=0.
\end{equation*}
\end{corollary}

Finally to consider the decomposition in terms of the projection mapping
onto the tangent space as in Section 2.3, we modify the definition of $%
\Gamma :$%
\begin{equation*}
\Gamma :\mathcal{L\rightarrow L},\text{ }A\mapsto P_{r}AP_{c},\text{ }%
P_{l_r}=I_{r}-\frac{1}{l_r}\mathbf{1}_{r}\mathbf{1}_{r}^{T},\text{ }%
P_{c}=I_{c}-\frac{1}{l_c}\mathbf{1}_{c}\mathbf{1}_{c}^{T}.
\end{equation*}%
and define $\mathbf{\Gamma }:\mathcal{L}^{2}\mathcal{\rightarrow L}^{2}$ by%
\begin{equation*}
(A,B)\mapsto \mathbb{P}(A,B)\mathbb{P}:\mathbb{=}%
\begin{pmatrix}
P_{r} & O \\ 
O & P_{c}%
\end{pmatrix}%
\begin{pmatrix}
O & A \\ 
B^{T} & O%
\end{pmatrix}%
\begin{pmatrix}
P_{r} & O \\ 
O & P_{c}%
\end{pmatrix}%
\,.
\end{equation*}%
As in symmetric games (Proposition \ref{prop:char_ker}), we obtain the
following characterizations for $\ker (\mathbf{\Gamma })$ and range $(%
\mathbf{\Gamma })$:

\begin{proposition}
We have 
\begin{eqnarray}
\ker(\mathbf{\Gamma}) &=& \mathrm{span}\left( \{(E_{\eta }^{(i)},O)\}_{i\neq
1}\cup \{(E_{\gamma }^{(i)},O)\}_{i}\cup \{(O,E_{\eta }^{(i)})\}_{i}\cup
\{(O,E_{\gamma }^{(i)})\}_{i\neq 1}\right)  \notag \\
\mathrm{range}(\mathbf{\Gamma}) \,&=&\, \mathrm{span} \left(\{(E_{\kappa
}^{(ij)},O)\}_{i\geq 1, j\geq 1}\cup \{O,E_{\kappa }^{(ij)}\}_{i\geq 1,j\geq
1}\right)  \notag
\end{eqnarray}
\end{proposition}

Clearly results similar to Proposition \ref{prop-space}, and Theorem \ref%
{cor:decomp} hold for $\mathcal{L}^{2}$ and the subspaces $\mathcal{\bar{M}}%
, $ $\mathcal{\bar{M}}^{\perp },$ $\mathcal{\bar{N}},\mathcal{\bar{N}}%
^{\perp },\ker (\mathbf{\Gamma })$, and range$(\mathbf{\Gamma })$. To
understand the relationship between the decompositions of symmetric games
and bi-matrix games, note that the set of two player symmetric games
corresponds to the set of all bi-matrix games with $l=l_{r}=l_{c}$
satisfying $A=B^{T}.$ Thus in this case, 
\begin{equation*}
(A,B)\text{ is a symmetric game if }A=B^{T}.
\end{equation*}%
To avoid confusion, we denote by $\mathcal{L}_{sym}$ the set of all
symmetric games as a subspace of $\mathcal{L}^{2}$ and write $[A]=(A,A^{T})$%
. Consider the following example:%
\begin{eqnarray*}
&&\,[E_{\kappa }^{(12)}-E_{\kappa }^{(21)}]=(E_{\kappa }^{(12)},-E_{\kappa
}^{(12)})-(E_{\kappa }^{(21)},-E_{\kappa }^{(21)}) \\
&=&%
\begin{tabular}{|l|l|l|}
\hline
0,0 & -1,1 & 1,-1 \\ \hline
0,0 & 1,-1 & -1,1 \\ \hline
0,0 & 0,0 & 0,0 \\ \hline
\end{tabular}%
-%
\begin{tabular}{|l|l|l|}
\hline
0,0 & 0,0 & 0,0 \\ \hline
-1,1 & 1,-1 & 0,0 \\ \hline
1,-1 & -1,1 & 0,0 \\ \hline
\end{tabular}%
=%
\begin{tabular}{|l|l|l|}
\hline
0,0 & -1,1 & 1,-1 \\ \hline
1,-1 & 0,0 & -1,1 \\ \hline
-1,1 & 1,-1 & 0,0 \\ \hline
\end{tabular}%
.
\end{eqnarray*}%
Thus \thinspace $\lbrack E_{\kappa }^{(12)}-E_{\kappa }^{(21)}]$ is the
Rock-Paper-Scissors game; this example shows how one can \textquotedblleft
symmetrize\textquotedblright\ the bi-matrix games to obtain the symmetric
version of them. More generally, we obtain the orthonormal bases of
anti-potential games and anti-zero-sum symmetric games in symmetric games by
restricting the bases of subspaces of bi-matrix games using the following
lemma.

\begin{lemma}
\label{lem-basis}Suppose that $\{(A^{(ij)},A^{(ij)})\}_{i,j\in \mathcal{I}%
_{1}}\cup \{(B^{(ij)},-B^{(ij)})\}_{i,j\in \mathcal{I}_{2}}\cup
\{(C^{(i)},O)\}_{i\in \mathcal{I}_{3}}\cup \{(O,(C^{(i)})^{T})\}_{i\in 
\mathcal{I}_{3}}$ form a basis for $K$, a subspace of $\mathcal{L}^{2}$ and $%
\{A^{(ij)}\}_{i,j}\cup \{B^{(ij)}\}_{i,j}\cup \{C^{(i)}\}_{i}$ are linearly
independent$.$ Then $\{[A^{(ij)}+A^{(ji)}]\}_{i,j\in \mathcal{I}_{1}\cap
\{j\geq i\}}\cup \{[B^{(ij)}-B^{(ji)}]\}_{i,j\in \mathcal{I}_{2}\cap
\{j>i\}}\cup \{[C^{(i)}]\}_{i\in \mathcal{I}_{3}}$ form a basis for $K\cap 
\mathcal{L}_{sym}.$
\end{lemma}

As an immediate consequence of the decomposition we obtain the alternative
proof for the following well-known characterization for potential and
zero-sum games \citep{Hofbauer98, Sandholm08}. Notice that a similar
characterization for the symmetric potential and zero-sum games is also
readily available.

\begin{proposition}
The following conditions are equivalent:\newline
(1) $(A,B)$ is a potential game (a zero-sum game, respectively)\newline
(2) $\ \mathbb{P}(A,B)\mathbb{P}$ is a symmetric $(l_{r}+l_{c})\times
(l_{r}+l_{c})$ matrix (an antisymmetric $(l_{r}+l_{c})\times (l_{r}+l_{c})$
matrix, respectively)\newline
(3) $(A,B)-(A,B)^{T}\in \ker (\mathbf{\Gamma })$ $\ ((A,B)+(A,B)^{T}\in \ker
(\mathbf{\Gamma })$, respectively.)
\end{proposition}

\begin{proof}
For a given $(A,B),$ using range($\mathbf{\Gamma })=\mathcal{M}^{\perp
}\oplus \mathcal{N}^{\perp }$ (Proposition \ref{prop-space}) we have%
\begin{equation*}
\mathbb{P}(A,B)\mathbb{P}=(V,V)+(N,-N)\text{ for some }V\text{ and }N\in 
\mathcal{L}.
\end{equation*}%
Since $(V,V)$ is a $(l_{r}+l_{c})\times (l_{r}+l_{c})$ symmetric matrix and $%
\mathbb{(}N,-N)$ is a $(l_{r}+l_{c})\times (l_{r}+l_{c})$ anti-symmetric, so 
$(1)$ $\Leftrightarrow (2).$ For $(2)$ $\Leftrightarrow (3),$ we first note
that $(A,B)^{T}=(B,A).$ Thus $(A\pm B,B\pm A)\in \ker (\mathbf{\Gamma }),$
if and only if $\mathbb{P}(A\pm B,B\pm A)\mathbb{P=}$ $O$, $\ $if and only
if $\mathbb{P}(A,B)\mathbb{P=\pm P}(B,A)\mathbb{P}$, if and only if $\mathbb{%
P}(A,B)\mathbb{P=\pm (P}(A,B)\mathbb{P)}^{T}.$
\end{proof}

\subsection{Decompositions of $n$-player normal form games.}

In this section we will briefly discuss how to generalize the decomposition
to the case of $n-$player normal form games. We provide more detailed
discussion in the Appendix. For the simplicity of exposition, we suppose
that all $n-$players have the same strategy set $S.$ We denote by $\mathcal{L%
}_{n}$ the set of all $n$ player games, by $\mathcal{S}$ the set of all
strategy profiles and by $\mathcal{P}$ the set of all players. First note
that we have $\dim (\mathcal{L}_{n})=nl^{n}.$ We use a $l^{n}$ dimensional
tensor $A$ to denote a player's payoffs and thus a normal form game is given
by $(A_{p_{1}},A_{p_{2}},\cdots ,A_{p_{n}})$ for $p_{l}\in \mathcal{P}.$ We
introduce an inner product $\left\langle {}\right\rangle _{\mathcal{L}_{n}}$
in $\mathcal{L}_{n\text{ }}:$%
\begin{equation*}
\left\langle (A_{p_{1}},\cdots ,A_{p_{n}}),(B_{p_{1}},\cdots
,B_{p_{n}})\right\rangle _{\mathcal{L}_{n}}=\sum_{i=1,\cdots n}\left\langle
A_{p_{i}},B_{p_{i}}\right\rangle _{\mathcal{L}}\,,
\end{equation*}%
\ where 
\begin{equation*}
\left\langle A,B\right\rangle _{\mathcal{L}}=\sum_{(i_{p_{1}},\cdots
,i_{p_{n}})\in S}a_{i_{p_{1}},\cdots ,i_{p_{n}}}b_{i_{p_{1}},\cdots
,i_{p_{n}}}.
\end{equation*}%
Similarly we denote by $\mathcal{M}_{n}$ the subspace of all potential
games. We have the following recursive formula for the dimension of $%
\mathcal{M}_{n}.$

\begin{proposition}
\label{prop:dim-anti-pot}We have $\dim (\mathcal{M}_{n+1})^{\perp
}=(l-1)^{2}nl^{n-1}+\dim (\mathcal{M}_{n})^{\perp }.$
\end{proposition}

\begin{proof}
First note that $\dim (\mathcal{M}_{n})=l^{n}-1+nl^{n-1}.$ and thus 
\begin{eqnarray*}
\dim (\mathcal{M}_{n+1})^{\perp } &=&(n+1)l^{n+1}-l^{n+1}-(n+1)l^{n}+1 \\
&=&(l-1)^{2}(nl^{n-1}+(n-1)l^{n-2}+\cdots +2l+1) \\
&=&(l-1)^{2}nl^{n-1}+\dim (\mathcal{M}_{n})^{\perp }.
\end{eqnarray*}
\end{proof}

\begin{figure}[tb]
\centering\includegraphics[scale=0.6]{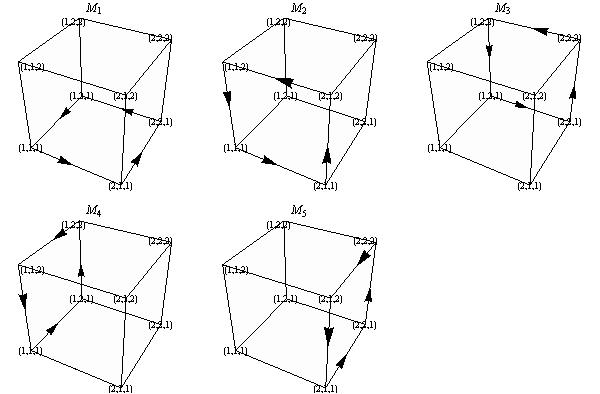}
\caption{\textbf{A basis for three-player anti-potential games. }Each vertex
in each cube represents the strategy profile and the arrows show the
deviation motivations based on the payoffs from the game.}
\label{fig:np}
\end{figure}

The recursive relation in Proposition \ref{prop:dim-anti-pot} shows that a
basis for $(\mathcal{M}_{n+1})^{\perp }$ can be obtained from the existing
basis of $(\mathcal{M}_{n})^{\perp }$ by adding $(l-1)^{2}nl^{n-1}$
additional elements. To illustrate this, we consider two strategy three
player games. From 
\begin{equation*}
\mathcal{M}_{2}=\mathrm{span}(%
\begin{tabular}{|l|l|}
\hline
-1,1 & 1,-1 \\ \hline
1,-1 & -1,1 \\ \hline
\end{tabular}%
),
\end{equation*}%
we expand this basis bi-matrix to obtain an element of the basis set for $%
\mathcal{M}_{3}$ by making player $3$ as a null player (See the first cubic
in Figure \ref{fig:np}) . That is,%
\begin{equation*}
M_{1}=%
\begin{tabular}{|l|l||l|l|}
\hline
-1,1,0 & 1,-1,0 & 0,0,0 & 0,0,0 \\ \hline
1,-1,0 & -1,1,0 & 0,0,0 & 0,0,0 \\ \hline
\end{tabular}%
.
\end{equation*}%
Now we imagine that one of existing players, player 1 and player 2, is
matched with player 3 to play the Matching Pennies game. Then since the null
player, either player 1 or player 2, can choose one strategy from the two
strategies, there are four possible situations in which two players play the
Matching Pennies game and one player plays the null player (See Figure \ref%
{fig:np}). Thus we obtain the following basis games.%
\begin{eqnarray*}
M_{2} &=&%
\begin{tabular}{|l|l||l|l|}
\hline
-1,0,1 & 0,0,0 & 1,0,-1 & 0,0,0 \\ \hline
1,0,-1 & 0,0,0 & -1,0,1 & 0,0,0 \\ \hline
\end{tabular}%
,\text{ }M_{3}=%
\begin{tabular}{|l|l||l|l|}
\hline
0,0,0 & -1,0,1 & 0,0,0 & 1,0,-1 \\ \hline
0,0,0 & 1,0,-1 & 0,0,0 & -1,0,1 \\ \hline
\end{tabular}
\\
M_{4} &=&%
\begin{tabular}{|l|l||l|l|}
\hline
0,-1,1 & 0,1,-1 & 0,1,-1 & 0,-1,1 \\ \hline
0,0,0 & 0,0,0 & 0,0,0 & 0,0,0 \\ \hline
\end{tabular}%
,\text{ }M_{5}=%
\begin{tabular}{|l|l||l|l|}
\hline
0,0,0 & 0,0,0 & 0,0,0 & 0,0,0 \\ \hline
0,-1,1 & 0,1,-1 & 0,1,-1 & 0,-1,1 \\ \hline
\end{tabular}%
\end{eqnarray*}

It is easy to see that $M_{1},\cdots ,M_{5}$ are independent and belong to $%
\mathcal{M}_{3}.$ Thus $\left\{ M_{1},\cdots ,M_{5}\right\} $ form a basis
for $\mathcal{M}_{3}.$ Here we verify Proposition \ref{prop:dim-anti-pot} as
follows:%
\begin{equation*}
\dim (\mathcal{M}_{3})^{\perp }=(2-1)^{2}2\times 2^{2-1}+\dim (\mathcal{M}%
_{2})^{\perp }.
\end{equation*}%
Note that 
\begin{equation*}
M_{6}=%
\begin{tabular}{|l|l||l|l|}
\hline
0,0,0 & 0,0,0 & -1,1,0 & 1,-1,0 \\ \hline
0,0,0 & 0,0,0 & 1,-1,0 & -1,1,0 \\ \hline
\end{tabular}%
\end{equation*}%
can be obtained by taking $M_{1}-(M_{2}-M_{3}-M_{4}+M_{5}).$ Next we
characterize the subspace of all zero-sum games. We call a game $\left(
A_{p_{1}},A_{p_{2}},\cdots ,A_{p_{n}}\right) $ is an exact zero-sum game if 
\begin{equation*}
(A_{p_{1}})_{(i_{p_{1}},\cdots ,i_{p_{n}})}+\cdots
+(A_{p_{n}})_{(i_{p_{1}},\cdots ,i_{p_{n}})}=0\text{ \ for all }%
(i_{p_{1}},\cdots ,i_{p_{n}})\in \mathcal{S}.
\end{equation*}%
The following lemma reveals the structure of the subspace of all zero-sum
games.

\begin{lemma}
\label{lem-zero}$A=\left( A_{p_{1}},A_{p_{2}},\cdots ,A_{p_{n}}\right) $ is
an exact zero-sum game if and only if $\ A$ can be written as a finite sum
of tensors $Z$'s of the form:%
\begin{equation*}
Z=(O,\cdots ,O,Z_{p_{i}},O,\cdots ,O,-Z_{p_{i}},O,\cdots ).
\end{equation*}
\end{lemma}

\begin{proof}
\textquotedblleft If part\textquotedblright\ is trivial. For
\textquotedblleft only if part\textquotedblright , we decompose $A$ first
into $l^{n}$ tensors whose $(i_{p_{1}},\cdots ,i_{p_{n}})$th element is the
same as $((A_{p_{1}})_{(i_{p_{1}},\cdots ,i_{p_{n}})},\cdots
,(A_{p_{n}})_{(i_{p_{1}},\cdots ,i_{p_{n}})})$ and other elements are all $0$%
's. $\ $Then since $((A_{p_{1}})_{(i_{p_{1}},\cdots ,i_{p_{n}})},\cdots
,(A_{p_{n}})_{(i_{p_{1}},\cdots ,i_{p_{n}})})\in T\Delta _{n}$ and $\left\{
(1,-1,0,\cdots ,0),(1,0,-1,\cdots ,0),\text{ }\cdots ,(1,0,0,\cdots
,-1)\right\} $ form a basis for $T\Delta _{n},$ we have the desired
representation.
\end{proof}

Then we can define the subspace of zero-sum games and obtain the following
proposition.

\begin{proposition}
\label{prop:dim-anti-zero}We have $\dim (\mathcal{N}^{\perp })=(l-1)^{p}$
\end{proposition}

From this discussion, we obtain the decompositions of potential games and
anti-potential games and the decompositions of zero-sum games and
anti-zero-sum games as in Section 2.1-2.3.

\subsection{Examples of Decompositions}

\begin{figure}[tb]
\centering
\includegraphics[scale=0.4]{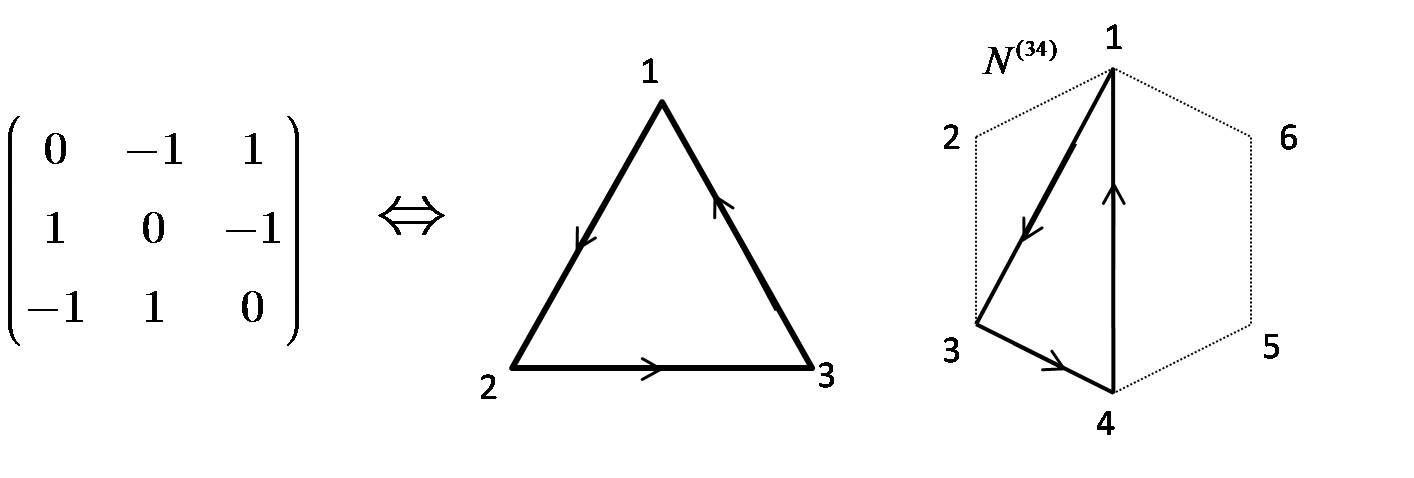}
\caption{\textbf{Representation of the Rock-paper-scissors games} 
{\protect\footnotesize {}}}
\label{fig:represent}
\end{figure}

%From Section 2.1-2.3, we see that a game $(A,B)$ $\in \mathcal{L}^{2}$ can
%be \textit{uniquely} decomposed into (i) a representative of equivalent
%classes of a potential game and an anti-potential game$,$ (ii) a
%representative of equivalent classes of a zero-sum game and an anti-zero-sum
%game, or (iii) an anti-potential part, an anti-zero-sum part, and an part
%belonging to $\ker (\mathbf{\Gamma })$. 
Because of the simple structure of basis games in the subspace of
anti-potential games we can associate a class of anti-potential games with a
set of graphs. To explain this we focus on symmetric games. First observe
that all basis elements in $\mathcal{M}^{\perp }$, $N^{(ij)}$ have payoffs
consisting $0$, $1$, and $-1$. Thus we can assign a binary relation to $%
(i,j) $: for given $A,$ $i\succ j$ if $a(i,j)=1$ ($i$ is better than $j$)$,$ 
$i\prec \,j$ \ if $a(i,j)=-1$ ($i$ is worse than $j$)$,$ and $i\sim j$ if $%
a(i,j)=0$ ($i$ is as good as $j).$ Since every anti-potential game is
anti-symmetric, the relation is symmetric; i.e., $i\succ j$ \ if and only if 
$j\prec i\,.$ Therefore we can represent a given basis element of
anti-potential games in a diagram as in Figure \ref{fig:represent}.

For games with cyclic symmetry \citep[][p.173]{Hofbauer98} we have the
following decomposition.

\begin{figure}[tb]
\centering
\includegraphics[scale=0.3]{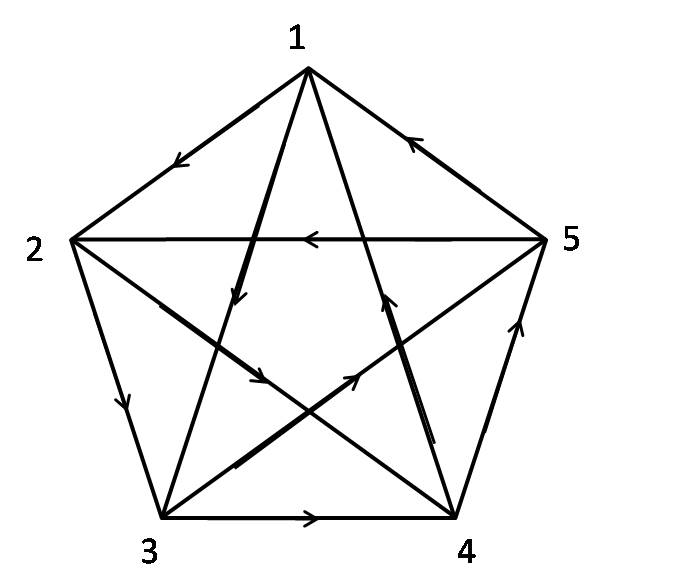}
\caption{\textbf{Games with cyclic symmetry.} {\protect\footnotesize {}}}
\label{fig:symmetry}
\end{figure}

\begingroup\setstretch{0.85}\scalefont{0.85}%
\begin{eqnarray*}
\begin{pmatrix}
0 & a_{1} & a_{2} & a_{3} & a_{4} \\ 
a_{4} & 0 & a_{1} & a_{2} & a_{3} \\ 
a_{3} & a_{4} & 0 & a_{1} & a_{2} \\ 
a_{2} & a_{3} & a_{4} & 0 & a_{1} \\ 
a_{1} & a_{2} & a_{3} & a_{4} & 0%
\end{pmatrix}
&\sim &\frac{1}{2}\underset{\mathcal{M}}{\underbrace{%
\begin{pmatrix}
0 & a_{1}+a_{4} & a_{2}+a_{3} & a_{2}+a_{3} & a_{1}+a_{4} \\ 
a_{1}+a_{4} & 0 & a_{1}+a_{4} & a_{2}+a_{3} & a_{2}+a_{3} \\ 
a_{2}+a_{3} & a_{1}+a_{4} & 0 & a_{1}+a_{4} & a_{2}+a_{3} \\ 
a_{2}+a_{3} & a_{2}+a_{3} & a_{1}+a_{4} & 0 & a_{1}+a_{4} \\ 
a_{1}+a_{4} & a_{2}+a_{3} & a_{2}+a_{3} & a_{1}+a_{4} & 0%
\end{pmatrix}%
}} \\
+ &&\frac{1}{2}\underset{\mathcal{M}^{\perp }}{\underbrace{%
\begin{pmatrix}
0 & a_{1}-a_{4} & a_{2}-a_{3} & -a_{2}+a_{3} & -a_{1}+a_{4} \\ 
-a_{1}+a_{4} & 0 & a_{1}-a_{4} & a_{2}-a_{3} & -a_{2}+a_{3} \\ 
-a_{2}+a_{3} & -a_{1}+a_{4} & 0 & a_{1}-a_{4} & a_{2}-a_{3} \\ 
a_{2}-a_{3} & -a_{2}+a_{3} & -a_{1}+a_{4} & 0 & a_{1}-a_{4} \\ 
a_{1}-a_{4} & a_{2}-a_{3} & -a_{2}+a_{3} & -a_{1}+a_{4} & 0%
\end{pmatrix}%
}}
\end{eqnarray*}%
\endgroup If $a_{1}-a_{4}=a_{2}-a_{3}-1,$ then the anti-potential part of
game can be represented in Figure \ref{fig:symmetry}.

In case of two-strategy bi-matrix coordination games, we have the following
decomposition of a two-strategy. 
\begin{eqnarray*}
&&%
\begin{tabular}{|l|l|}
\hline
$a,b$ & $0,0$ \\ \hline
$0,0$ & $c,d$ \\ \hline
\end{tabular}%
\sim \frac{1}{2}\underset{\ker (\mathbf{\Gamma })}{\underbrace{%
\begin{tabular}{|c|c|}
\hline
$0,0$ & $0,-b+d$ \\ \hline
$-a+c,0$ & $-a+c,-b+d$ \\ \hline
\end{tabular}%
}}+\frac{1}{8}(a+b+c+d)\underset{\mathcal{\bar{N}}^{\perp }}{\underbrace{%
\begin{tabular}{|c|c|}
\hline
$1,1$ & $-1,-1$ \\ \hline
$-1,-1$ & $1,1$ \\ \hline
\end{tabular}%
}} \\
&&+\frac{1}{8}(-a+b-c+d)\underset{\mathcal{\bar{M}}^{\perp }}{\underbrace{%
\begin{tabular}{|l|l|}
\hline
$-1,1$ & $1,-1$ \\ \hline
$1,-1$ & $-1,1$ \\ \hline
\end{tabular}%
}}
\end{eqnarray*}%
Therefore, a two-strategy coordination game is a potential game if and only
if $-a+b-c+d=0$ and a zero-sum game if and only if $a+b+c+d=0.$ In other
words, the coefficients of the anti-potential game and the anti-zero-sum
game corresponds to the condition for payoffs in four-cycle criteria as in
Corollary \ref{cor:bi_pot} and \ref{cor:bi_zero}.

\section{Applications of Decompositions}

\subsection{Decompositions and Stable Games}

In this section, we provide a characterization of stable games 
\citep[for
properties of stable games see][]{Hofbauer09}. 
%We recall the definition of stable games in terms of matrix notations in order
%to facilitate the applications of our decompositions. 
A symmetric game $[A]$ is a stable game if $\left\langle
y-x,A(y-x)\right\rangle _{\mathbb{R}_{l}}\leq 0$ \ for all $x,y\in \Delta
_{l}.$ A bi-matrix game $(A,B)$ is a stable game if $\left\langle
y-x,(A,B)(y-x)\right\rangle _{\mathbb{R}_{l_{r}+l_{c}}}\leq 0$ \ for all $%
x,y\in \Delta _{l_{r}}\times \Delta _{l_{c}}.$ A null-stable game is a
stable for which equality holds instead of an inequality.

Note that since $[A]=(A,A^{T}),$ the condition for a symmetric game to be
stable can be written as 
\begin{equation}
\left\langle y-x,(A,A^{T})(y-x)\right\rangle _{\mathbb{R}_{l+l}}\leq 0\text{
\ for all }x,y\in \{(p,q)\in \Delta _{l}\times \Delta _{l}:p=q\} \,.
\label{def-stable3}
\end{equation}%
By comparing this to the condition for bi-matrix games to be stable (with $%
l_{r}=l_{c}=l$) we see that the inequality in (\ref{def-stable3}) holds for
a smaller subset of $\mathbb{R}^{2l}.$ This opens the possibility that more
stable games arise in symmetric games. 
% even though symmetric games belong to the special
%class of bi-matrix games. 
Using the projection operator $P$ defined in Section 2 we see that a
symmetric game $A$ is a stable game if and only if $\left\langle
x,PAPx\right\rangle \leq 0$ \ for all \ $x\in \mathbb{R}_{l}$ and a
bi-matrix game $(A,B)$ is a stable game if and only if $\left\langle x,%
\mathbb{P}(A,B)\mathbb{P}x\right\rangle \leq 0$ for all $x\in \mathbb{R}%
_{l_{r}+l_{c}}$ \citep[][Theorem 2.1]{Hofbauer09}$.$

We first characterize stable symmetric matrix games. To do this we define a
function $V_{A\text{ }}$ for a given symmetric game $A$, which will play an
important role in characterizing stable games: $V_{A}(x):=\frac{1}{2}%
\left\langle x,Ax\right\rangle .$ Then using the decomposition, we obtain
the following representation of $V_{A}.$

\begin{proposition}
\label{prop:strict-stable}Suppose that $A \in \mathcal{L}$. Then there
exists a symmetric matrix $S$ with $S\mathbf{1}=0$ and a column vector a
vector $c$ such that, for any $x\in \Delta $ and any $z\in T\Delta $ 
\begin{equation*}
V_{A}(x)=\frac{1}{2}\left\langle x,Sx\right\rangle +\left\langle
x,c\right\rangle ,\text{ }\ V_{A}(z)=\frac{1}{2}\left\langle
z,Sz\right\rangle .\text{ }
\end{equation*}
Moreover there exists an orthonormal basis $\{v_1, \cdots, v_{l-1}\}$ of $%
T\Delta$ (in particular $\langle v_i, \mathbf{{1} \rangle =0}$) such that $%
S=\sum_{i=1}^{l-1}\lambda_{i}S_{i}$ where $S_{i}$ is the orthogonal
projection onto the eigenspace spanned by $v_{i}$.
\end{proposition}

\begin{proof}
Let $A\in \mathcal{L}= \mathcal{N}^{\perp }\oplus \ker (\Gamma )\oplus 
\mathcal{M}^{\perp }$ and thus we can write%
\begin{equation*}
A=S+c_{1}\mathbf{1}^{T}+\mathbf{1}c_{2}^{T}+N
\end{equation*}%
where $S$ is symmetric with $S\mathbf{1}=0$ and $N$ is anti-symmetric with $N%
\mathbf{1}=0$. Thus for any $x \in \mathbb{R}^l$ we have 
\begin{equation*}
V_{A}(x)=\frac{1}{2}\left\langle x,Sx\right\rangle +\frac{1}{2}\left\langle
x,(c_{1}\mathbf{1}^{T}+\mathbf{1}c_{2}^{T})x\right\rangle
\end{equation*}%
For $x \in \Delta$ we have $\left\langle x,c_{1}\mathbf{1}^{T}x\right\rangle
=\sum_{i}x_{i}\left\langle x,c_{1}\right\rangle =\left\langle
x,c_{1}\right\rangle $ and $\left\langle x,\mathbf{1}c_{2}^{T}x\right\rangle
=\sum_{i}x_{i}\left\langle x,c_{2}\right\rangle =\left\langle
x,c_{2}\right\rangle$, and thus 
\begin{equation*}
\frac{1}{2}\left\langle x,(c_{1}\mathbf{1}^{T}+\mathbf{1}c_{2}^{T})x\right%
\rangle =\left\langle x,c\right\rangle \text{ }\ \ \ \text{where }%
c=c_{1}+c_{2}.
\end{equation*}%
Note further that $S=PSP$ and $\left\langle z,c_{1}\mathbf{1}%
^{T}z\right\rangle =\left\langle z,\mathbf{1}c_{2}^{T}z\right\rangle =0$ for 
$z \in T\Delta$. Since $S$ is symmetric, all eigenvectors are orthogonal and
since $\mathbf{1}$ is an eigenvector with the corresponding eigenvalue $%
\lambda =0,$ all other eigenvectors belong to $T\Delta $ and the
representation of $S$ follows from the spectral theorem.
\end{proof}

To characterize the stable games using Proposition \ref{prop:strict-stable},
we let $A\in \mathcal{L}$ and $z\in T\Delta .$ Then 
\begin{equation*}
V_{A}(z)=\frac{1}{2}\left\langle z,Sz\right\rangle =\frac{1}{2}\left\langle
\sum_{i}\xi _{i}v_{i},S\sum_{i}\xi _{i}v_{i}\right\rangle =\frac{1}{2}%
\sum_{i}\xi _{i}^{2}\lambda _{i}
\end{equation*}%
where $v_{i}$ is orthonormal basis for $T\Delta $ consisting of eigenvectors
of $S.$ Thus $A$ is null-stable iff $\lambda _{i}=0$ for all $i$. 
%Then since $%
%S$ is a symmetric matrix, $S=0$ if and only if all its eigenvalues are 0.
Therefore $A$ is a null-stable game if and only if $A\in \mathcal{N}.$ We
put this fact as Proposition \ref{prop:sym_null_stable} of which another
direct proof is presented in the Appendix. Similarly note that $V_{A}(z)$ $%
<0 $ for all $z\neq 0$ if and only if $\lambda _{i}<0$ for all $i.$ \ Thus a
game is a strict stable game if and only if the eigenvalues for $S,$ except
the one corresponding to $\mathbf{1},$ are all negative.

\begin{proposition}
\label{prop:sym_null_stable}$\left\langle x,PAPx\right\rangle =0$ for all $%
x\in \mathbb{R}^{l}$ if and only if $A\in \mathcal{N}\mathbf{.}$
\end{proposition}

As is well-known,\textsl{\ the Hawk-Dove game }provides the simplest
possible strictly stable game, and from the equivalence we find%
\begin{equation*}
(z_{1},z_{2})%
\begin{pmatrix}
-1 & 1 \\ 
1 & -1%
\end{pmatrix}%
\begin{pmatrix}
z_{1} \\ 
z_{2}%
\end{pmatrix}%
=-(z_{1}-z_{2})^{2}<0,\ \text{for }(z_{1},z_{2})\neq (0,0).
\end{equation*}%
This observation can be generalized via the basis of the subspace of
anti-zero-sum games $\mathcal{N}^{\perp }$ $.$

\begin{corollary}[$l-$strategy strictly stable games]
\label{cor:l_st_strict_stable}Suppose that 
\begin{equation*}
\text{ }A\in \{\sum_{j>i}\alpha ^{(ij)}K^{(ij)}:\alpha ^{(ij)}>0\}+\ker (%
\mathbf{\Gamma })+\mathcal{M}^{\perp }.
\end{equation*}%
Then $A$ is a strict stable game.
\end{corollary}

\begin{proof}
Recall that $[A]$ is a strict stable game if $\left\langle z,Az\right\rangle
<0$ for all $z\in T\Delta $ such that $z\neq \mathbf{0}.$Let $A\in S.$ Then
we have $\left\langle z,Az\right\rangle =-\sum_{j>i}\alpha
^{(ij)}(z_{i}-z_{j})^{2}\leq 0.$Now suppose that $-\sum_{j>i}\alpha
^{(ij)}(z_{i}-z_{j})^{2}=0.$ Then we have $z_{i}-z_{j}=0$ for all $j>i.$
Since $z\in T\Delta ,$ this implies that $z=\mathbf{0}.$
\end{proof}

In case of three-strategy games, we can strengthen Corollary \ref%
{cor:l_st_strict_stable} so as to characterize three-strategy strict stable
games completely, since the computation in three-strategy case is less
demanding.

\begin{corollary}[three-strategy strictly stable games]
\label{cor:3_st_strict_stable}A $three-$strategy symmetric game $A$ is
strictly stable if and only if 
\begin{equation*}
A\in \left\{ 
\begin{pmatrix}
-a-b & a & b \\ 
a & -a-c & c \\ 
b & c & -b-c%
\end{pmatrix}%
:4a+b+c>0,\text{ }ab+bc+ca>0\right\} +\ker (\Gamma )+\mathcal{M}^{\perp }.
\end{equation*}
\end{corollary}

First we note that when $l=3$ in Corollary \ref{cor:l_st_strict_stable} the
condition for strictly stable games is a special case of Corollary \ref%
{cor:3_st_strict_stable} by the choices of $a,b>0$ and $c=0.$ As another
important special case of Corollary \ref{cor:3_st_strict_stable}, consider
game $B$ given by%
\begin{equation*}
B=%
\begin{pmatrix}
0 & \beta _{12} & \beta _{13} \\ 
\beta _{12} & 0 & \beta _{23} \\ 
\beta _{13} & \beta _{23} & 0%
\end{pmatrix}%
.
\end{equation*}%
First note that $B$ is a potential game, so there is no anti-potential part
of $B.$ Thus $B$ can be decomposed into 
\begin{eqnarray*}
B &=&%
\begin{pmatrix}
-a-b & a & b \\ 
a & -a-c & c \\ 
b & c & -b-c%
\end{pmatrix}%
+\underset{\in \text{ }\ker (\Gamma )}{\underbrace{C}}\text{ and } \\
a &=&\frac{1}{9}(5\beta _{12}-\beta _{13}-\beta _{23}),b=\frac{1}{9}(-\beta
_{12}+5\beta _{13}-2\beta _{23}),c=\frac{1}{9}(-\beta _{12}-\beta
_{13}+5\beta _{23}).
\end{eqnarray*}%
Then the conditions in Corollary \ref{cor:3_st_strict_stable} imply 
\begin{equation}
\beta _{12}>0\text{ \ \ and }(\beta _{12}+\beta _{23}+\beta
_{13})^{2}>2(\beta _{12}^{2}+\beta _{23}^{2}+\beta _{13}^{2}).
\label{con_stab}
\end{equation}%
Recall that the generalized Rock-Paper-Scissors game can be decomposed as
follows:%
\begin{equation*}
\begin{pmatrix}
0 & -l & w \\ 
w & 0 & -l \\ 
-l & w & 0%
\end{pmatrix}%
\sim \frac{1}{2}(w-l)%
\begin{pmatrix}
0 & 1 & 1 \\ 
1 & 0 & 1 \\ 
1 & 1 & 0%
\end{pmatrix}%
+\frac{1}{2}(w+l)%
\begin{pmatrix}
0 & -1 & 1 \\ 
1 & 0 & -1 \\ 
-1 & 1 & 0%
\end{pmatrix}%
.
\end{equation*}%
We see that the case when $\beta _{12}=\beta _{23}=\beta _{13},$ $\beta
_{12}>0$ satisfies conditions in (\ref{con_stab}), so using Corollary \ref%
{cor:3_st_strict_stable} we conclude that the generalized
Rock-Paper-Scissors game is strictly stable if and only if $w>l$ \ (See the
discussion in \citep[See the discussion in][]{Hofbauer09}). In the next
section we will provide another useful parametrization of three-strategy
anti-zero-sum games.

Next we characterize the bi-matrix stable games. First we recall that for $J$
given by%
\begin{equation*}
J:=%
\begin{pmatrix}
O & A \\ 
B^{T} & O%
\end{pmatrix}%
,
\end{equation*}%
the characteristic polynomial $p(\lambda )=\det (J-\lambda I)$ satisfies $%
p(\lambda )=(-1)^{l_{r}+l_{c}}p(-\lambda ).$ Hence if $\lambda $ is an
eigenvalue, then $-\lambda $ is also an eigenvalue. For a given bi-matrix
game $(A,B),$ we can write $(A,B)\sim (V,V)+(C,D)+(N,-N)$ where $(C,D)\in
\ker (\mathbf{\Gamma })$. Thus, $\mathbb{P}(A,B)\mathbb{P=}$ $\mathbb{P}(V,V)%
\mathbb{P}$. $\ $So if $(A,B)$ is a stable game, all its eigenvalues must
have the same sign and, thus, they must be all zeros. Hence every stable
bi-matrix game is always null-stable \citep[][Theorem2.1]{Hofbauer09}. Then,
as the similar argument as Proposition \ref{prop:strict-stable} shows, every
null-stable bi-matrix game is a zero-sum game. \ As a result, we provide the
complete characterization of the set of all stable bi-matrix games; the set
of all stable bi-matrix games is the set of all zero-sum games. Proposition %
\ref{prop:asym_null_stable} can be proved via either the straightforward
extension of Proposition \ref{prop:sym_null_stable} or the direct use of the
basis of decompositions. We provide the direct proof in the Appendix.

\begin{proposition}
\label{prop:asym_null_stable}$\left\langle w,\mathbb{P}(A,B)\mathbb{P}%
w\right\rangle =0$ for all $w\in \mathbb{R}_{l_{r}+l_{c}}$ if and only if $%
(A,B)\in \mathcal{\bar{N}}.$
\end{proposition}

\subsection{Decomposition and Deterministic Dynamics}

Evolutionary dynamics based on the normal form games have been extensively
examined and their important properties are closely related to the
underlying games; for example, potential games yield the gradient like
replicator dynamics \citep{Hofbauer98}. Moreover the replicator dynamics are
linear with respect to the underlying game matrix (or matrices), so our
decompositions naturally induce decompositions at the level of vector
fields. We will consider the replicator dynamics given by%
\begin{equation}
\text{One population: \ \ }\dot{x}_{i}=x_{i}((Ax)_{i}-x^{T}Ax)\text{ for all 
}i\text{ \ \ }  \label{rep}
\end{equation}%
\begin{equation*}
\text{Two population: }x_{i}=x_{i}((Ay)_{i}-x^{T}Ay),\text{ \ }\dot{y}%
_{j}=y_{j}((B^{T}x)_{j}-y^{T}B^{T}x)
\end{equation*}

When we have $A\sim S+G+N,$ where $S\in \mathcal{N}^{\perp },$ $G\in \ker
(\Gamma ),$ $N\in \mathcal{M}^{\perp },$ the replicator dynamics can also be
decomposed in three parts. $\ $First note that if $G=\sum_{i}\eta
_{i}E_{\eta }^{(i)},$ then $(Gx)_{i}=\eta _{i}$ and $\left\langle
x,Gx\right\rangle =\sum_{l\neq 1}\eta _{l}x_{l},$ so the vector field for
the replicator dynamics induced by $G$ is given by%
\begin{equation*}
x_{i}(\eta _{i}-\sum_{l\neq 1}\eta _{l}x_{l})
\end{equation*}%
and the system monotonically moves towards the dominating strategy state. \
Also when $x^{T}Nx=0$ for $N\in \mathcal{M}^{\perp }.$ Thus, the replicator
ordinary differential equation for the matrix $A$ can be decomposed into 
\begin{equation*}
\begin{tabular}{llllll}
$f_{i}(x)\sim $ & $\underset{\text{potential part}}{\underbrace{%
x_{i}((Sx)_{i}-x^{T}Sx)}}$ & + & $\underset{\text{monotonic part}}{%
\underbrace{x_{i}(\eta _{i}-\sum_{l\neq 1}\eta _{l}x_{l})}}$ & + & $\underset%
{\text{conservative part}}{\underbrace{x_{i}Nx}}$ \\ 
& \includegraphics[scale=0.5]{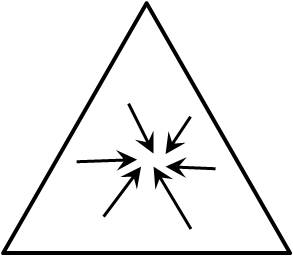} &  & %
\includegraphics[scale=0.5]{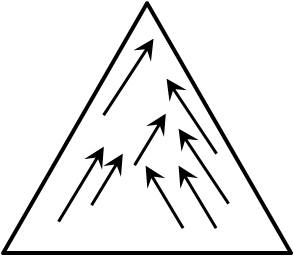} &  & %
\includegraphics[scale=0.5]{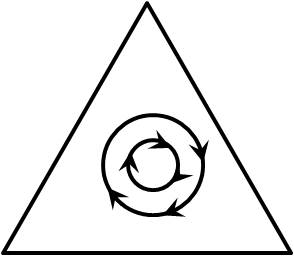}%
\end{tabular}%
\end{equation*}%
This decomposition of the vector field of the replicator ordinary
differential equations coincides with the known Hodge decomposition which
plays an important role in understanding the underlying dynamics (See %
\citet{Abraham88} and equation (1) in \citet{Tong03}).

We recall that a function $H$: $D\rightarrow \mathbb{R}$ is an integral of (%
\ref{rep}) on a region $D$ if $H$ is continuous differentiable and $H(x(t))$
is constant along the solution of (\ref{rep})$;$i.e., $LH(x(t)):=\left%
\langle \nabla H(x(t)),f(x(t))\right\rangle =0$ for a solution $x(t).$ The
orbits of a conservative system must therefore lie on level curves of the
integral $H.$ A system $(\ref{rep})$ is said to be conservative if it has an
integral $H.$ \ We again recall that a function $V$: $D\rightarrow \mathbb{R}
$ is a strict Lyapunov function for $C\subset D$ if $V$ is continuous that
achieves its minimum at $C$ is non-increasing along the solutions and is
decreasing outside of $C;$ i.e., $LV(x):=\left\langle \nabla
V(x),f(x))\right\rangle \leq 0$ for $x$~in $D$ and $LV(x)<0$ for $x\notin C.$

It is well-known that the replicator dynamic for the Rock-Paper-Scissors
games is conservative and volume-preserving, the dynamics of the Matching
Pennies games can be transformed to Hamiltonian systems by change in
velocity of solutions, and all the bi-matrix games preserve volume up to
change in velocity of solutions \citep{Hofbauer98}. As Proposition \ref%
{prop:con-vol} shows, the class of anti-potential games provides the
dynamics which are volume-preserving without involving the change of time.

\begin{proposition}
\label{prop:con-vol}(1) Suppose that $[A]$ is an anti-potential game. Then (%
\ref{rep}) is conservative and volume-preserving.\newline
(2) Suppose $(A,B)$ $\in range(\mathbf{\Gamma }).$ Then \ $(A,B)$ is
conservative.\newline
(3) Suppose that $(A,B)$ is an anti-potential game and $l_{c}=l_{r}$. Then $%
(A,B)$ is volume preserving.
\end{proposition}

In the generalized Rock-Paper-Scissors game, it is easy to check that when $%
b>a$, $\ H(x):=\sum_{i}\log (x_{i})$ is a strict Lyapunov function for $%
\mathbf{(}\frac{1}{3},\frac{1}{3},\frac{1}{3}).$ Our decompositions show
that this observation generalizes to the bigger class of games that have the
similar structure to the generalized Rock-Paper-Scissors game.

\begin{proposition}
Suppose 
\begin{equation*}
A\in \{\sum_{j>i}\alpha ^{(ij)}K:\alpha ^{(ij)}>0\}+\mathcal{M}^{\perp }%
\text{ }.
\end{equation*}%
Then, $H(x):=\sum_{i}\log (x_{i})$ is a strict Lyapunov function for $\frac{1%
}{n}\mathbf{1}$. And thus a unique NE $\frac{1}{n}\mathbf{1}$ is
evolutionarily stable.
\end{proposition}

\begin{proof}
Let $A=S+N,$ where $S\in \{\sum_{j>i}\alpha ^{(ij)}K:\alpha ^{(ij)}>0\}$ and 
$N\in \mathcal{M}^{\perp }.$ Note that for $x\neq \frac{1}{n}\mathbf{1,}$ we
have%
\begin{eqnarray*}
LH &=&\sum_{i}((Ax)_{i}-x^{T}Ax)=\sum_{i}(Sx)_{i}-\left\langle
x,Sx\right\rangle +\sum_{i}(Nx)_{i} \\
&=&-\left\langle x,Sx\right\rangle =-\left\langle x,PSPx\right\rangle
=-\left\langle z,Sz\right\rangle >0
\end{eqnarray*}
\end{proof}

Next we explain how to obtain the game which has a pure strategy ESS and an
interior asymptotically stable NE (called the Zeeman game) using the
decomposition. We consider a game $A\in \mathcal{N}^{\perp }\oplus \mathcal{M%
}^{\perp }.$ Then since $A\mathbf{1=0,}$ $\frac{1}{n}\mathbf{1}$ is a Nash
equilibrium. From the previous discussion, the anti-zero-sum part $S$ of $A$
is completely determined by its eigenvalues and (orthonormal) eigenvectors.
Recall that $S$ always has an eigenvector $\mathbf{1}$ with the
corresponding eigenvalue $0$ and other eigenvectors lie in the tangent
space. Thus when the number of strategies is three, any two eigenvectors in
the tangent space can be obtain by rotating given reference orthogonal
eigenvectors around the axis $(1,1,1). $ First we denote the matrix for the
Rock-Paper-Scissors game by $N: $%
\begin{equation*}
N:=%
\begin{pmatrix}
0 & -1 & 1 \\ 
1 & 0 & -1 \\ 
-1 & 1 & 0%
\end{pmatrix}%
\end{equation*}%
Next, to express this parameterization of $S$ we define the rotation matrix $%
R $ which rotates a given vector in $\mathbb{R}^{3}$ around the axis $%
(1,1,1),$ as follows:%
\begin{equation}
R=I-P+(\cos \theta I+\sin \theta \frac{1}{\sqrt{3}}N)P.  \label{eq:rotation}
\end{equation}%
To explain the meaning of $R,$ we first recall that the rotation matrix in $%
\mathbb{R}^{2}$ acts as follows:%
\begin{equation*}
\begin{pmatrix}
\cos \theta & -\sin \theta \\ 
\sin \theta & \cos \theta%
\end{pmatrix}%
x=\cos \theta Ix+\sin \theta 
\begin{pmatrix}
0 & -1 \\ 
1 & 0%
\end{pmatrix}%
x.
\end{equation*}%
Thus the rotation matrix map $x$ to the linear combination of $x$ itself and
a vector orthogonal to $x,$ and the coefficients of the combination are
parameterized by an angle. Now note that $\left\langle Nx,x\right\rangle =0$
for all $x.$ Thus when $z\in T\Delta ,$%
\begin{equation*}
Rz=\cos \theta Iz+\sin \theta \frac{1}{\sqrt{3}}Nz
\end{equation*}%
and since $Nz$ is orthogonal to $z,$ $R$ acts in the same way as the
rotation in two-dimension. Also clearly $R\mathbf{1=0.}$ When $x\in \mathbb{R%
}^{3},$ $x$ can be uniquely written as $x=(I-P)x+Px$ and $R$ rotates the
part belonging to $range(P).$ Thus, $R$ has the representation in (\ref%
{eq:rotation}). $\ $Using the rotation matrix $R,$ we can write a
three-strategy game $A$ as follows$:$ 
\begin{eqnarray*}
A &=&R%
\begin{pmatrix}
\alpha +\frac{\beta }{3} & -\frac{2\beta }{3} & -\alpha +\frac{\beta }{3} \\ 
-\frac{2\beta }{3} & \frac{4\beta }{3} & -\frac{2\beta }{3} \\ 
-\alpha +\frac{\beta }{3} & -\frac{2\beta }{3} & \alpha +\frac{\beta }{3}%
\end{pmatrix}%
R^{-1}+\eta 
\begin{pmatrix}
0 & 1 & -1 \\ 
-1 & 0 & 1 \\ 
1 & -1 & 0%
\end{pmatrix}%
, \\
&&\text{where }%
\begin{pmatrix}
\alpha +\frac{\beta }{3} & -\frac{2\beta }{3} & -\alpha +\frac{\beta }{3} \\ 
-\frac{2\beta }{3} & \frac{4\beta }{3} & -\frac{2\beta }{3} \\ 
-\alpha +\frac{\beta }{3} & -\frac{2\beta }{3} & \alpha +\frac{\beta }{3}%
\end{pmatrix}%
=%
\begin{pmatrix}
1 & 1 & 1 \\ 
1 & 0 & -2 \\ 
1 & -1 & 1%
\end{pmatrix}%
\begin{pmatrix}
0 & 0 & 0 \\ 
0 & 2\alpha & 0 \\ 
0 & 0 & 2\beta%
\end{pmatrix}%
\begin{pmatrix}
1 & 1 & 1 \\ 
1 & 0 & -2 \\ 
1 & -1 & 1%
\end{pmatrix}%
^{-1}
\end{eqnarray*}%
Then the matrix $A$ has the characteristic polynomial $\phi
(t)=t(t^{2}-2(\alpha +\beta )t+4\alpha \beta +3\eta ^{2})$, so it has the
eigenvalues $0,$ $\alpha +\beta \pm \sqrt{(\alpha -\beta )^{2}-3\eta ^{2}}$
and the eigenvector $\mathbf{1}$ corresponding to the eigenvalue $0$. Note
the eigenvalues for $A$ do not depend on the choice of $\theta .$ We can
also verify this as follows. From $RN=NR,$ we have%
\begin{equation*}
A=REDE^{-1}R^{-1}+\eta N=REDE^{-1}R^{-1}+\eta RNR^{-1}=R(EDE^{-1}+\eta
N)R^{-1},
\end{equation*}%
where $E$ denotes the matrix whose columns consist of orthogonal
eigenvectors and $D$ denotes the diagonal matrix which has $0,$ $2\alpha ,$%
and 2$\beta $ on the diagonal. Since $R(EDE^{-1}+\eta N)R^{-1}$ has the same
eigenvalues as $EDE^{-1}+\eta N,$ eigenvalues of $A$ do not depend on the
particular choice of $\theta .$

Since $\frac{\partial }{\partial x_{i}}(x^{T}Ax)=(Ax)_{i}+(A^{T}x)_{i},\,$
by differentiating (\ref{rep}), we find that 
\begin{equation}
\frac{\partial f_{i}(x)}{\partial x_{j}}=x_{i}(a_{ij}-(Ax)_{j}-(A^{T}x)_{j})%
\text{ \ \ for }j\neq i\text{, }\frac{\partial f_{i}(x)}{\partial x_{i}}%
=((Ax)_{i}-xA^{T}x)+x_{i}(a_{ii}-(Ax)_{i}).\text{ \ \ }  \label{eq:Jac}
\end{equation}%
So if we evaluate the expressions in (\ref{eq:Jac}) at $x=\frac{1}{n}\mathbf{%
1,}$ from $A\mathbf{1}=\mathbf{0}$, $A^{T}\mathbf{1=0,}$and $%
(Ax)_{i}-xA^{T}x=0,$ we find the following Jacobian matrix%
\begin{equation*}
\left. \frac{\partial f_{i}(x)}{\partial x_{j}}\right\vert _{x=\frac{1}{n}%
\mathbf{1}}=\frac{1}{n}a_{ij}.
\end{equation*}%
Thus the eigenvalues for the linearized system around $\frac{1}{n}\mathbf{1}$
are the same as the ones for $A$ up to a scalar multiple $\frac{1}{n}$. Also
we note that if $(\alpha -\beta )^{2}<3\eta ^{2}$, then two non-zero
eigenvalues are complex and in this case real parts of eigenvalues are
negative (zero, positive, resp.) if and only if $\alpha +\beta <0$ ($\alpha
+\beta =0,\alpha +\beta >0, $resp.$).$ Now we set $\theta =0.$ Then 
\begin{equation*}
A=\frac{1}{3}%
\begin{pmatrix}
3\alpha +\beta & -2\beta +3\eta & -3\alpha +\beta -3\eta \\ 
-2\beta -3\eta & 4\beta & -2\beta +3\eta \\ 
-3\alpha +\beta +3\eta & -2\beta -3\eta & 3\alpha +\beta%
\end{pmatrix}%
\end{equation*}%
so it is easy to see that if $-(\alpha +\beta )<\eta <2\alpha ,$ then
strategy 1 is a strict Nash equilibrium, hence an evolutionary stable
strategy. Thus we obtain the following characterization of Zeeman games.

\begin{proposition}
Suppose that $-(\alpha +\beta )<\eta <2\alpha ,$ and $(\alpha -\beta
)^{2}<3\eta ^{2}.$ Then strategy 1 is an ESS and the interior fixed point is
a sink (center, source, resp.) if $\alpha +\beta <0$ ($\alpha +\beta
=0,\alpha +\beta >0,$resp.$)$.
\end{proposition}

In Figure \ref{fig:rotation} we show how the vector field of the system
changes when $\theta $ varies. \ To find a four-strategy Zeeman game we
consider the following matrix using the similar idea:

\begin{figure}[tb]
\centering\includegraphics[scale=0.4]{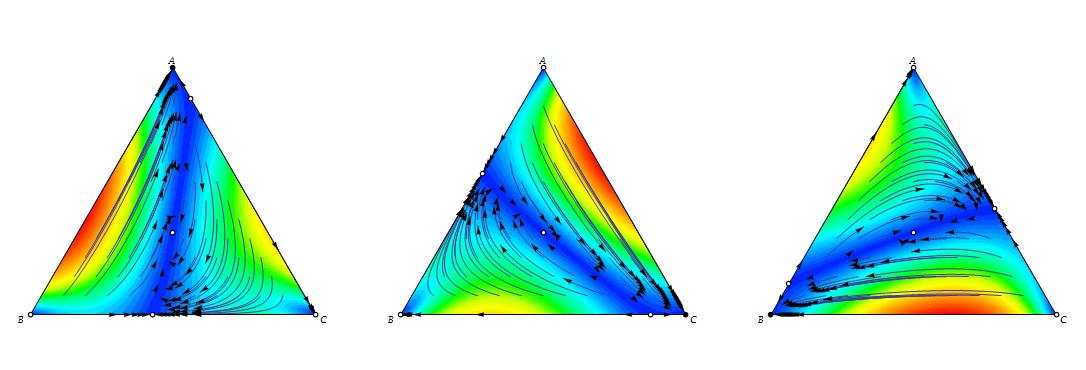}
\caption{\textbf{Rotation of Eigenvectors in the Zeeman Games. }Figures are
drawn using Dynamo: W. H. Sandholm, E. Dokumaci, and F. Franchetti (2010).
Dynamo: Diagrams for Evolutionary Game Dynamics, version 0.2.5. }
\label{fig:rotation}
\end{figure}

%\begingroup\setstretch{1}\scalefont{0.9}%
\begin{equation*}
A=%
\begin{pmatrix}
1 & 1 & 1 & 1 \\ 
1 & 0 & 0 & -3 \\ 
1 & 0 & -2 & 1 \\ 
1 & -1 & 1 & 1%
\end{pmatrix}%
\begin{pmatrix}
0 & 0 & 0 & 0 \\ 
0 & \alpha & 0 & 0 \\ 
0 & 0 & \beta & 0 \\ 
0 & 0 & 0 & \gamma%
\end{pmatrix}%
\begin{pmatrix}
1 & 1 & 1 & 1 \\ 
1 & 0 & 0 & -3 \\ 
1 & 0 & -2 & 1 \\ 
1 & -1 & 1 & 1%
\end{pmatrix}%
^{-1}+\eta 
\begin{pmatrix}
0 & 1 & 0 & -1 \\ 
-1 & 0 & 1 & 0 \\ 
0 & -1 & 0 & 1 \\ 
1 & 0 & -1 & 0%
\end{pmatrix}%
\end{equation*}%
%
%
%
%
%
%
%
%
%
%
%
%
%
%
%
%
%
%
%
%
%
%
%
%\endgroup
Then, it is easy to see that if $-\gamma <\eta <\gamma $ $\ $and\ $\gamma
>0, $ strategy 2 become a strict Nash equilibrium, so an ESS. The
characteristic polynomial for $A$ is 
\begin{equation*}
\phi (t)=t(t^{3}-(\alpha +\beta +\gamma )t^{2}+(\alpha \beta +\beta \gamma
+\gamma \alpha +4\eta ^{2})t-\alpha \beta \gamma -\frac{1}{3}(6\alpha
+2\beta +4\gamma )\eta ^{2}).
\end{equation*}%
Thus from the Routh-Hurwitz criterion\citep[For example see][]{Murray89}, we
see that eigenvalues $\lambda $ for $A$ all have negative real parts (except 
$0$ eigenvalue) if and only if 
\begin{eqnarray*}
&&\alpha +\beta +\gamma <0,\text{ }\alpha \beta +\beta \gamma +\gamma \alpha
+4\eta ^{2}>0,\text{ }3\alpha \beta \gamma +(6\alpha +2\beta +4\gamma )\eta
^{2}<0, \\
&&\text{ }\alpha \beta \gamma +\frac{1}{3}(6\alpha +2\beta +4\gamma )\eta
^{2}>(\alpha +\beta +\gamma )(\alpha \beta +\beta \gamma +\gamma \alpha
+4\eta ^{2}).
\end{eqnarray*}%
Using these conditions we exhibit a four-strategy Zeeman game in Figure \ref%
{fig:4x4-zeeman}. 
\begin{figure}[tb]
\centering 
\includegraphics[scale=0.4]{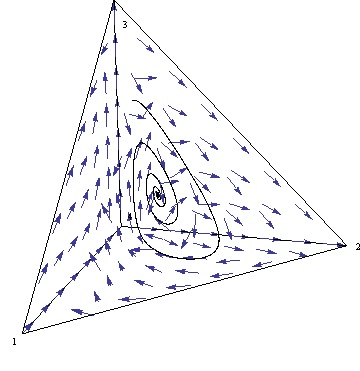}
\caption{\textbf{4-strategy Zeeman game. } {\protect\footnotesize {$\protect%
\alpha :-2.5,$ $\protect\beta :-2.5,$ $\protect\gamma :2,~\protect\eta :1.9.$
}}}
\label{fig:4x4-zeeman}
\end{figure}

\section{Conclusion}

We have developed several decomposition methods for two player normal form
games and discussed the extension to the general normal form games. Using
decompositions, we characterize (1) the subspaces of potential games and
their orthogonal complements, anti-potential games, (2) the subspaces of
zero-sum games and their orthogonal complements, anti-zero-sum games, and
(3) the subspaces of both potential and zero-sum games and their orthogonal
complements. Notably, the subspaces of anti-potential games consist of
special games, the Rock-Paper-Scissors games in the case of the symmetric
games and the Matching Pennies games in the case of the bi-matrix games. We
have explained how the previous known criterion for the potential games can
be viewed from the perspective of decompositions and provided a new cycle
criterion for symmetric zero-sum games.

We have discussed the various applications of the decompositions, including
(1) the analysis of the generalized Rock-Paper-Scissors games, (2) the
characterization of the stable games, (3) the decomposition of the vector
field and the construction of Lyapunov functions in evolutionary dynamics,
and so on. These decompositions turn out to be useful in the analysis of the
stochastic dynamics of the evolutionary games; these applications will be
discussed elsewhere.

\appendix\setstretch{1} \scalefont{0.8}\newpage

\section{Appendix}

\subsection{Properties of inner products}

First we observe that

\begin{enumerate}
\item $(A,B)^{T}=(B,A)$

\item $(A,B)$ is symmetric in $\mathcal{L}^{2}$ if $A=B$

\item $(A,B)$ is anti-symmetric in $\mathcal{L}^{2}$ if $A=-B$

\item $(A,B)$ is a symmetric game $\ $if $l_{r}=l_{c}$ and $A=B^{T}$
\end{enumerate}

So, a bi-matrix symmetric game is not necessarily a symmetric matrix in $%
\mathcal{L}^{2}.$ We endow $\mathcal{L}^{2}$ with an inner product $<,>_{%
\mathcal{L}^{2}}$ defined by:%
\begin{equation*}
\left\langle (A,B),(C,D)\right\rangle _{\mathcal{L}^{2}}:=tr((A,B)^{T}(C,D))
\end{equation*}%
We provide some properties of $\left\langle \text{ }\right\rangle _{\mathcal{%
L}^{2}}$ and $\left\langle \text{ }\right\rangle _{\mathcal{L}}$.

\begin{lemma}
\label{lem-inner}For $(l_{r}\times l_{c})$ matrices $A,B,C,D$ $,$ we have

\begin{enumerate}
\item[(1)] $\left\langle (A,B),(C,D)\right\rangle _{\mathcal{L}%
^{2}}=\left\langle A,C\right\rangle _{\mathcal{L}}+\left\langle
B,D\right\rangle _{\mathcal{L}}$

\item[(2)] $\left\langle SA,B\right\rangle _{\mathcal{L}}=\left\langle
A,SB\right\rangle _{\mathcal{L}}$ for a symmetric $(l_{r}\times l_{r})$
matrix $S$

\item[(3)] $\left\langle (A,A),(B,-B)\right\rangle _{\mathcal{L}^{2}}=0$

\item[(4)] For $\mathbf{c}\in \mathbb{R}^{l_{r}}$ and $A$ such that $A%
\mathbf{1}_{l_{c}}=\mathbf{0},$ $\left\langle A,\mathbf{c1}%
_{l_{c}}^{T}\right\rangle _{\mathcal{L}}=0.$

\item[(5)] For $\mathbf{c}\in \mathbb{R}^{l_{c}}$ and $A$ such that $\mathbf{%
1}_{l_{r}}^{T}A=\mathbf{0},$ $\left\langle A,\mathbf{1}_{l_{r}}\mathbf{c}%
^{T}\right\rangle _{\mathcal{L}}=0.$
\end{enumerate}
\end{lemma}

\begin{proof}
(1) and (2) are obvious. (3) follows from 
\begin{equation*}
\left\langle (A,A),(B,-B)\right\rangle _{\mathcal{L}^{2}}=\left\langle
A,B\right\rangle _{\mathcal{L}}-\left\langle A,B\right\rangle _{\mathcal{L}%
}=0.
\end{equation*}%
(4) follows from 
\begin{equation*}
\left\langle A,\mathbf{c1}_{l_{c}}^{T}\right\rangle =tr(\mathbf{1}_{l_{c}}%
\mathbf{c}^{T}A)=tr(\mathbf{c}^{T}A\mathbf{1}_{l_{c}})=0
\end{equation*}%
by the commutativity of trace and (5) follows from 
\begin{equation*}
\left\langle A,\mathbf{1}_{l_{r}}\mathbf{c}^{T}\right\rangle _{\mathcal{L}%
}=tr(\mathbf{c1}_{l_{r}}^{T}A)=0
\end{equation*}
\end{proof}

\subsection{Proof of Proposition \protect\ref{prop:char_ker}}

\begin{proof}
(1) We first show that 
\begin{equation*}
\ker \Gamma =\mathrm{span}\{E_{\eta }^{(1)},\cdots ,E_{\eta
}^{(l)},E_{\gamma }^{(1)},\cdots ,E_{\gamma }^{(l)}\}
\end{equation*}%
Note that $PE_{\gamma }^{(j)}=O$ for all $j.$ Then $E_{\eta
}^{(i)}P=(P(E_{\eta }^{(i)})^{T})^{T}=O$ for all $i.$ Thus we have $\mathrm{%
span}$ $\{E_{\eta }^{(1)},\cdots ,E_{\eta }^{(l)},E_{\gamma }^{(1)},\cdots
,E_{\gamma }^{(l)}\}\subset \ker \Gamma $. Conversely, let $A$ such that $%
\Gamma (A)=O.$ Since%
\begin{equation*}
PAP=A-\frac{1}{l}\mathbf{11}^{T}A-\frac{1}{l}A\mathbf{11}^{T}+\frac{1}{l^{2}}%
\mathbf{11}^{T}A\mathbf{11}^{T},
\end{equation*}%
we have%
\begin{equation*}
A=\frac{1}{l}\mathbf{11}^{T}A+\frac{1}{l}A\mathbf{11}^{T}-\frac{1}{l^{2}}%
\mathbf{11}^{T}A\mathbf{11}^{T}
\end{equation*}%
Then note the following properties of $\mathbf{11}^{T}:$%
\begin{equation*}
\mathbf{11}^{T}A=(\sum_{k}a_{k1}\mathbf{1}:\sum_{k}a_{k2}\mathbf{1}:\cdots
:\sum_{k}a_{kl}\mathbf{1})
\end{equation*}%
i.e., the left action of $\mathbf{11}^{T}$ on $A$ turns $A$ into a matrix
with the same elements in each column. Since $A\mathbf{11}^{T}=(\mathbf{11}%
^{T}A^{T})^{T\text{ }},$ the right action of $\mathbf{11}^{T}$ on $A$ turns $%
A$ into a matrix with same elements in each column. Also it is easy to see
that $\mathbf{11}^{T}A\mathbf{11}^{T}=\sum_{k}\sum_{m}a_{km}\mathbf{11}^{T}.$
Thus $A$ can be written as%
\begin{equation*}
A=\sum_{i}(\sum_{k}a_{ki})E_{\gamma }^{(i)}+\sum_{j}(\sum_{k}a_{jk})E_{\eta
}^{(j)}+(\sum_{k}\sum_{m}a_{km})\sum_{j}E_{\gamma }^{(j)}
\end{equation*}%
So $A\in \mathrm{span}\{E_{\eta }^{(1)},\cdots ,E_{\eta }^{(l)},E_{\gamma
}^{(1)},\cdots ,E_{\gamma }^{(l)}\}.$ Thus $\ker \Gamma =\mathrm{span}%
\{E_{\eta }^{(1)},\cdots ,E_{\eta }^{(l)},$ $E_{\gamma }^{(1)},\cdots
,E_{\gamma }^{(l)}\}.$ Next note that%
\begin{equation*}
\sum_{i}E_{\eta }^{(i)}=\sum_{j}E_{\gamma }^{(j)},\text{ \ so }E_{\gamma
}^{(1)}=-\sum_{j\neq 1}E_{\gamma }^{(j)}-\sum_{i}E_{\eta }^{(i)},
\end{equation*}%
thus 
\begin{equation*}
\mathrm{span}\{E_{\eta }^{(1)},\cdots ,E_{\eta }^{(l)},E_{\gamma
}^{(1)},\cdots ,E_{\gamma }^{(l)}\}=\mathrm{span}\{E_{\eta }^{(2)},\cdots
,E_{\eta }^{(l)},E_{\gamma }^{(1)},\cdots ,E_{\gamma }^{(l)}\}.
\end{equation*}%
To show the linear independence among $\{E_{\eta }^{(2)},\cdots ,E_{\eta
}^{(l)},E_{\gamma }^{(1)},\cdots ,E_{\gamma }^{(l)}\},$ consider the linear
combination of these matrices:%
\begin{equation*}
O=\sum_{i\neq 2}\eta _{i}E_{\eta }^{(2)}+\sum_{j}\gamma _{j}E_{j}^{(2)}.
\end{equation*}%
Then since $E_{\eta }^{(1)}$ does not appear in the linear combination, we
have $\gamma _{j}=0$ for all $j$ and this implies $\eta _{i}=0$ for $i\neq
2. $ \newline
(2) Note because of $\mathbf{1}E_{\kappa }^{(ij)}=\mathbf{0}$ and $E_{\kappa
}^{(ij)}\mathbf{1}=\mathbf{0},\Gamma (E_{\kappa }^{(ij)})=PE_{\kappa
}^{(ij)}P=E_{\kappa }^{(ij)}.$ So, 
\begin{equation*}
E_{\kappa }^{(ij)}\subset \text{range(}\Gamma \text{) \ \ for all }i,j\geq 2
\end{equation*}%
and it is easy to see that $E_{\kappa }^{(ij)}$ are linearly independent.
Finally by $\left\vert \{E_{\kappa }^{(ij)}\}_{ij}\right\vert =(l-1)^{2}$
and since dim$($range($\Gamma ))=(l-1)^{2},$ $\{E_{\kappa }^{(ij)}\}_{ij}$
is a basis for range$(\Gamma )$
\end{proof}

\subsection{Proof of Proposition \protect\ref{prop-space}}

\begin{proof}
First we show that $\ker (\Gamma )=\mathcal{M\cap N}$. \ Observe that for $%
i\geq 2$%
\begin{equation*}
(E_{\eta }^{(i)}+E_{\gamma }^{(i)})^{T}=(E_{\eta }^{(i)})^{T}+(E_{\gamma
}^{(i)})^{T}=E_{\gamma }^{(i)}+E_{\eta }^{(i)}
\end{equation*}%
Thus $(E_{\eta }^{(i)}+E_{\gamma }^{(i)})$ is symmetric and $(E_{\eta
}^{(i)}+E_{\gamma }^{(i)})_{11}=0,$ so $E_{\eta }^{(i)}=(E_{\eta
}^{(i)}+E_{\gamma }^{(i)})^{T}-E_{\gamma }^{(i)}\in \mathcal{M}.$ Also%
\begin{equation*}
(E_{\eta }^{(i)}-E_{\gamma }^{(i)})^{T}=(E_{\eta }^{(i)})^{T}-(E_{\gamma
}^{(i)})^{T}=-(E_{\eta }^{(i)}-E_{\gamma }^{(i)}),
\end{equation*}%
so $(E_{\eta }^{(i)}-E_{\gamma }^{(i)})$ is anti-symmetric and $E_{\eta
}^{(i)}=-(E_{\eta }^{(i)}-E_{\gamma }^{(i)})^{T}+E_{\gamma }^{(i)}\in 
\mathcal{N}.$ Therefore $\ker \Gamma \subset $ $\mathcal{M}\cap \mathcal{N}.$
Conversely let $\ A\in \mathcal{M}\cap \mathcal{N}.$ Then 
\begin{equation*}
A=S+\mathbf{1c}_{1}^{T}\text{ and }A=B+\mathbf{1c}_{2}^{T}\text{ for a
symmetric }S\text{ and anti-symmetric }B.
\end{equation*}%
Thus $B+\mathbf{1c}_{2}^{T}-\mathbf{1c}_{1}^{T}$ = $B^{T}+\mathbf{c}_{2}%
\mathbf{1}^{T}-\mathbf{c}_{1}\mathbf{1}^{T}$ and using the anti-symmetry of $%
B,$ we obtain%
\begin{equation*}
B=\frac{1}{2}(\mathbf{c}_{2}\mathbf{1}^{T}-\mathbf{c}_{1}\mathbf{1}^{T}+%
\mathbf{1c}_{2}^{T}-\mathbf{1c}_{1}^{T})
\end{equation*}%
and so 
\begin{equation*}
A=\frac{1}{2}(\mathbf{c}_{2}\mathbf{1}^{T}-\mathbf{c}_{1}\mathbf{1}^{T}+%
\mathbf{1c}_{2}^{T}-\mathbf{1c}_{1}^{T})+\mathbf{1c}_{2}^{T}\in \ker \Gamma .
\end{equation*}%
Next we show that range$(\Gamma )=\mathrm{span}(\mathcal{M}^{\perp }\cup 
\mathcal{N}^{\perp }).$ Then, we have 
\begin{eqnarray*}
\mathrm{span}(\mathcal{M}^{\perp }\cup \mathcal{N}^{\perp }) &=&\mathrm{span}%
(\{N^{(ij)}\}_{j>i\geq 2}\cup \{H^{(ij)}\}_{j>i\geq 2}\cup
\{K^{(ii)}\}_{i\geq 2}\} \\
&=&\mathrm{span}(\{K^{(ij)}\}_{j>i\geq 2}\cup \{K^{(ij)}\}_{i>j\geq 2}\cup
\{K^{(ii)}\}_{i\geq 2}\} \\
&=&\mathrm{span}(\{K^{(ij)}\}_{i\geq 2,j\geq 2}\}=\text{range}(\Gamma )
\end{eqnarray*}
\end{proof}

\subsection{Proof of Lemma \protect\ref{lem-basis}}

\begin{proof}
First we show that $\mathrm{span}(\{[A^{(ij)}+A^{(ji)}]\}_{i,j\in \mathcal{I}%
_{1}\cap \{j\geq i\}}\cup \{[B^{(ij)}-B^{(ji)}]\}_{i,j\in \mathcal{I}%
_{2}\cap \{j>i\}}\cup \{[C^{(i)}]\}_{i\in \mathcal{I}_{3}})=\mathcal{K\cap L}%
_{sym}.$ Obviously,%
\begin{eqnarray*}
\lbrack A^{(ij)}+A^{(ji)}]
&=&(A^{(ij)}+A^{(ji)},(A^{(ij)})^{T}+(A^{(ji)})^{T})=(A^{(ij)},A^{(ji)})+(A^{(ji)},A^{(ij)})
\\
&=&(A^{(ij)},A^{(ij)})+(A^{(ji)},A^{(ji)})\in \mathcal{K\cap L}_{sym}
\end{eqnarray*}%
Similarly we have $\{(B^{(ij)},-B^{(ij)})\}_{i,j}\in \mathcal{K\cap L}%
_{sym}. $ Also $%
[C^{(i)}]=(C^{(i)},(C^{(i)})^{T})=(C^{(i)},O)+(O,(C^{(i)})^{T})\in \mathcal{%
K\cap L}_{sym}.$ Conversely, let $(E,F)\in \mathcal{K\cap L}_{sym}.$ Then 
\begin{eqnarray*}
&&(E,F) \\
&=&\sum_{i,j\in \mathcal{I}_{1}}\kappa
_{(1)}^{(ij)}(A^{(ij)},A^{(ij)})+\sum_{i,j\in \mathcal{I}_{2}}\kappa
_{(2)}^{(ij)}(B^{(ij)},-B^{(ij)})+\sum_{i\in \mathcal{I}_{3}}\kappa
_{(3)}^{(i)}(C^{(i)},O)+\sum_{i\in \mathcal{I}_{3}}\kappa
_{(4)}^{(i)}(O,(C^{(i)})^{T}) \\
&=&(\sum_{i,j\in \mathcal{I}_{1}}\kappa _{(1)}^{(ij)}A^{(ij)}+\sum_{i,j\in 
\mathcal{I}_{2}}\kappa _{(2)}^{(ij)}B^{(ij)}+\sum_{i\in \mathcal{I}%
_{3}}\kappa _{(3)}^{(i)}C^{(i)}, \\
&&\sum_{i,j\in \mathcal{I}_{1}}\kappa _{(1)}^{(ij)}A^{(ij)}-\sum_{i,j\in 
\mathcal{I}_{2}}\kappa _{(2)}^{(ij)}B^{(ij)}+\sum_{i\in \mathcal{I}%
_{3}}\kappa _{(4)}^{(i)}(C^{(i)})^{T})
\end{eqnarray*}%
Since $E=F^{T},$ we have%
\begin{equation*}
\sum_{i,j\in \mathcal{I}_{1}}\kappa _{(1)}^{(ij)}A^{(ij)}+\sum_{i,j\in 
\mathcal{I}_{2}}\kappa _{(2)}^{(ij)}B^{(ij)}+\sum_{i\in \mathcal{I}%
_{3}}\kappa _{(3)}^{(i)}C^{(i)}=\sum_{i,j\in \mathcal{I}_{1}}\kappa
_{(1)}^{(ij)}A^{(ji)}-\sum_{i,j\in \mathcal{I}_{2}}\kappa
_{(2)}^{(ij)}B^{(ji)}+\sum_{i\in \mathcal{I}_{3}}\kappa _{(4)}^{(i)}C^{(i)}
\end{equation*}%
Thus we obtain%
\begin{equation}
\sum_{i,j\in \mathcal{I}_{1}}(\kappa _{(1)}^{(ij)}-\kappa
_{(1)}^{(ji)})A^{(ij)}+\sum_{i,j\in \mathcal{I}_{2}}(\kappa
_{(2)}^{(ij)}+\kappa _{(2)}^{(ji)})B^{(ij)}+\sum_{i\in \mathcal{I}%
_{3}}(\kappa _{(3)}^{(i)}-\kappa _{(4)}^{(i)})C^{(i)}=O  \label{eq:lem_1}
\end{equation}%
Then from the linear independency of $\{A^{(ij)}\}_{ij}\cup
\{B^{(ij)}\}_{ij}\cup \{C^{(i)}\}_{i}$ in $\mathcal{L},$ we conclude that 
\begin{equation*}
\kappa _{(1)}^{(ij)}=\kappa _{(1)}^{(ji)},\text{ }\kappa
_{(2)}^{(ij)}=-\kappa _{(2)}^{(ji)},\text{ and }\kappa _{(3)}^{(i)}=\kappa
_{(4)}^{(i)}\text{ for all }i,j
\end{equation*}%
Note that $\kappa _{(2)}^{(ii)}=0$ for all $i.$ Thus we have%
\begin{eqnarray*}
&&\sum_{i,j\in \mathcal{I}_{1}}\kappa _{(1)}^{(ij)}(A^{(ij)},A^{(ij)}) \\
&=&\sum_{\{j>i\}\cap \mathcal{I}_{1}}\kappa
_{(1)}^{(ij)}(A^{(ij)},A^{(ij)})+\sum_{\{j<i\}\cap \mathcal{I}_{1}}\kappa
_{(1)}^{(ij)}(A^{(ij)},A^{(ij)})+\sum_{\{i=j\}\cap \mathcal{I}_{1}}\kappa
_{(1)}^{(ij)}(A^{(ij)},A^{(ij)}) \\
&=&\sum_{\{j>i\}\cap \mathcal{I}_{1}}\kappa
_{(1)}^{(ij)}(A^{(ij)},A^{(ij)})+\sum_{\{j>i\}\cap \mathcal{I}_{1}}\kappa
_{(1)}^{(ji)}(A^{(ji)},A^{(ji)})+\sum_{\{(i,i)\}\cap \mathcal{I}_{1}}\kappa
_{(1)}^{(ii)}(A^{(ii)},A^{(ii)}) \\
&=&\sum_{\{j>i\}\cap \mathcal{I}_{1}}\kappa
_{(1)}^{(ij)}((A^{(ij)}+A^{(ji)},A^{(ij)}+A^{(ji)})+\sum_{\{(i,i)\}\cap 
\mathcal{I}_{1}}\frac{1}{2}\kappa
_{(1)}^{(ii)}(A^{(ii)}+A^{(ii)},A^{(ii)}+A^{(ii)}) \\
&=&\sum_{\{j>i\}\cap \mathcal{I}_{1}}\kappa
_{(1)}^{(ij)}((A^{(ij)}+A^{(ji)},A^{(ji)}+A^{(ij)})+\sum_{\{(i,i)\}\cap 
\mathcal{I}_{1}}\frac{1}{2}\kappa
_{(1)}^{(ii)}(A^{(ii)}+A^{(ii)},A^{(ii)}+A^{(ii)}) \\
&=&\sum_{\{j>i\}\cap \mathcal{I}_{1}}\kappa _{(1)}^{(ij)}\left[
A^{(ij)}+A^{(ji)}\right] +\sum_{\{(i,i)\}\cap \mathcal{I}_{1}}\frac{1}{2}%
\kappa _{(1)}^{(ii)}[A^{(ii)}+A^{(ii)}]
\end{eqnarray*}%
Similar manipulation yields 
\begin{eqnarray*}
&&\sum_{i,j\in \mathcal{I}_{2}}\kappa _{(2)}^{(ij)}(B^{(ij)},-B^{(ij)}) \\
&=&\sum_{\{j>i\}\cap \mathcal{I}_{2}}\kappa
_{(2)}^{(ij)}(B^{(ij)},-B^{(ij)})+\sum_{\{j<i\}\cap \mathcal{I}_{2}}\kappa
_{(2)}^{(ij)}(B^{(ij)},-B^{(ij)}) \\
&=&\sum_{\{j>i\}\cap \mathcal{I}_{2}}\kappa
_{(2)}^{(ij)}(B^{(ij)},-B^{(ij)})+\sum_{\{j>i\}\cap \mathcal{I}_{2}}\kappa
_{(2)}^{(ji)}(B^{(ji)},-B^{(ji)}) \\
&=&\sum_{\{j>i\}\cap \mathcal{I}_{2}}\kappa
_{(2)}^{(ij)}((B^{(ij)}-B^{(ji)},-B^{(ij)}+B^{(ji)})=\sum_{\{j>i\}\cap 
\mathcal{I}_{2}}\kappa _{(2)}^{(ij)}((B^{(ij)}-B^{(ji)},B^{(ji)}-B^{(ij)}) \\
&=&\sum_{\{j>i\}\cap \mathcal{I}_{2}}\kappa _{(2)}^{(ij)}\left[
B^{(ij)}-B^{(ji)}\right]
\end{eqnarray*}%
and finally%
\begin{equation*}
\sum_{i}\kappa _{(3)}^{(i)}(C^{(i)},O)+\sum_{i}\kappa
_{(4)}^{(i)}(O,(C^{(i)})^{T})=\sum_{i}\kappa
_{(3)}^{(i)}(C^{(i)},(C^{(i)})^{T})=\sum_{i}\kappa _{(3)}^{(i)}[C^{(i)}].
\end{equation*}%
Therefore, we have $\mathrm{span}(\{[A^{(ij)}+A^{(ji)}]\}_{i,j}\cup
\{[B^{(ij)}-B^{(ji)}]\}_{i,j}\cup \{[C^{(i)}]\}_{i})=\mathcal{K\cap L}%
_{sym}. $ Next we show that $\{[A^{(ij)}+A^{(ji)}]\}_{i,j}\cup
\{[B^{(ij)}-B^{(ji)}]\}_{i,j}\cup \{[C^{(i)}]\}_{i}$ are linearly
independent in $\mathcal{L}^{2}.$ Suppose that 
\begin{equation*}
\sum_{\{j\geq i\}\cap \mathcal{I}_{1}}\alpha
_{ij}[A^{(ij)}+A^{(ji)}]+\sum_{\{j>i\}\cap \mathcal{I}_{2}}\beta
_{ij}[B^{(ij)}-B^{(ji)}]+\sum_{\mathcal{I}_{3}}\gamma _{i}[C^{(i)}]=O\text{
in }\mathcal{L}^{2}
\end{equation*}%
Then we have 
\begin{equation*}
\sum_{\{j\geq i\}\cap \mathcal{I}_{1}}\alpha
_{ij}(A^{(ij)}+A^{(ji)})+\sum_{\{j>i\}\cap \mathcal{I}_{2}}\beta
_{ij}(B^{(ij)}-B^{(ji)})+\sum_{\mathcal{I}_{3}}\gamma _{i}(C^{(i)})=O\text{
in }\mathcal{L}
\end{equation*}%
and note that we have 
\begin{eqnarray*}
\sum_{\{j\geq i\}\cap \mathcal{I}_{1}}\alpha _{ij}(A^{(ij)}+A^{(ji)})
&=&\sum_{\{j>i\}\cap \mathcal{I}_{1}}\alpha _{ij}A^{(ij)}+\sum_{\{i>j\}\cap 
\mathcal{I}_{1}}\alpha _{ji}A^{(ij)}+\sum_{\{i=j\}\cap \mathcal{I}%
_{3}}\alpha _{ii}A^{(ii)} \\
\sum_{\{j>i\}\cap \mathcal{I}_{2}}\beta _{ij}(B^{(ij)}-B^{(ji)})
&=&\sum_{\{j>i\}\cap \mathcal{I}_{2}}\beta _{ij}B^{(ij)}-\sum_{\{i>j\}\cap 
\mathcal{I}_{2}}\beta _{ji}B^{(ij)}
\end{eqnarray*}%
and since $\left\{ A^{(ij)}\right\} _{i,j}\cup \left\{ B^{(ij)}\right\}
_{i,j}\cup \left\{ C^{(i)}\right\} _{i}$ are linearly independent in $%
\mathcal{L}$, we conclude that $\alpha _{ij}=0,\beta _{ij}=0,$ and $\gamma
_{i}=0\,\ $for all $i,j.$
\end{proof}

\subsection{Decomposition of $n-$player games}

We will denote by $S$, $S_{-p}\,,$ and $S_{-p\cup q}$ the set of all
strategy profiles, the set of all strategy profiles except player $p,$and
the set of all strategy profiles except player $p$ and $q$ $;$ i.e.,%
\begin{eqnarray*}
\mathcal{S} &:&=\{\left( i_{p_{1}},\cdots ,i_{p_{n}}\right)
:i_{p_{1}},\cdots ,i_{p_{n}}\in S\} \\
\mathcal{S}_{-q} &:&=\{\left( i_{p_{1}},\cdots ,\hat{\imath}_{q},\cdots
,i_{p_{n}}\right) :i_{p_{1}},\cdots ,i_{p_{n}}\in S\} \\
\mathcal{S}_{-q\cup r} &:&=\{\left( i_{p_{1}},\cdots ,\hat{\imath}%
_{q},\cdots ,\hat{\imath}_{r}\cdots ,i_{p_{n}}\right) :i_{p_{1}},\cdots
,i_{p_{n}}\in S\},
\end{eqnarray*}%
where $\hat{\imath}_{q}$ means that we omit the $q$th element. Then it is
easy to see that $\left\vert \mathcal{S}_{-p}\right\vert =l^{n-1}.$ Also for 
$\vec{i}_{-q\cup r}\in S_{-q\cup \gamma },$ 
\begin{equation*}
(A_{p_{1}})_{\vec{i}_{-q\cup r}}:=(A_{p_{1}})_{(i_{p_{1}},\cdots ,\hat{\imath%
}_{q},\cdots ,\hat{\imath}_{r},\cdots ,i_{p_{n})}}
\end{equation*}%
can be written as an $l\times l$ matrix and for $\vec{i}_{-q}\in
S_{-q},(A_{p_{1}})_{\vec{i}_{-q}}$ can be written as a $l\times 1$ vector.
We also write%
\begin{equation*}
\vec{i}_{-q}\subset \vec{j}\text{ \ if }(i_{p_{1}},\cdots ,k,\cdots
,i_{p_{n}})=(j_{p_{1}},\cdots ,j_{q},\cdots ,j_{p_{n}})\text{ \ for some }%
k\in S
\end{equation*}%
\ for $\vec{i}_{-q}\in S_{-q}$ and $\vec{j}\in S$.$\ $To define passive
games, we define a tensor $E_{\gamma }^{\vec{i}_{-q}}$ for $\vec{i}_{-q}\in
S_{-q}$ as follows:%
\begin{equation}
(E_{\gamma }^{\vec{i}_{-q}})_{\vec{i}_{-q}}=\mathbf{1}\text{ and }0\text{'s
in other positions}  \label{eq:pass}
\end{equation}%
where $1$ denotes a $l\times 1$ vector consisting of 1's. Then $E_{\gamma }^{%
\vec{i}_{-q}}$ in (\ref{eq:pass}) is an tensor that describes the payoffs of
player $q$ and under this payoffs, given other players' strategy profile $%
(i_{1},\cdots ,\hat{\imath}_{q},\cdots ,i_{n})$ for any choice of $q$
player's strategy, $q$ obtains payoff 1. Then similarly we set 
\begin{equation*}
\mathcal{I}=\mathrm{span}(\{(E_{\gamma }^{\vec{i}_{-p_{1}}},O,\cdots ,O)\}_{%
\vec{i}_{-p_{1}}\in \mathcal{S}_{-p_{1}}},\cdots ,\{(O,\cdots ,E_{\gamma }^{%
\vec{i}_{-p_{n}}})\}_{\vec{i}_{-p_{n}}\in \mathcal{S}_{-p_{n}}})
\end{equation*}%
where $O$ denotes a $l^{n}-$ dimensional zero tensor. Then $I$ is the set of
all passive games. We also define the following tensors: \ for $\vec{i}\in
S, $%
\begin{equation*}
(E_{\beta }^{\vec{i}})_{\vec{j}}=1\text{ if }\vec{i}=\vec{j}\text{ .}
\end{equation*}%
\ Then $E_{\beta }^{\vec{i}}$ is a tensor which has 1 at the position $\vec{i%
}$ and 0's at others. Then similarly we set $\mathcal{M}:=\mathrm{span}%
(\{(E_{\beta }^{\vec{i}},\cdots ,E_{\beta }^{\vec{i}})\}_{\vec{i}\in 
\mathcal{S}},I\}.$ Then we obtain Proposition \ref{prop:dim-anti-pot}. $\ $%
Next, using Lemma \ref{lem-zero}, we define the subspace of all zero-sum
games:%
\begin{equation*}
\mathcal{N}:=\mathrm{span}(\{(O,\cdots ,\underset{i\text{th}}{\underbrace{%
E_{\beta }^{\vec{i}}}},\cdots ,\underset{j\text{th}}{\underbrace{-E_{\beta
}^{\vec{i}}}},\cdots ,O)\}_{(i_{p_{1}},\cdots ,i_{p_{n}})\in \mathcal{S}%
,p_{i},p_{j}\in \mathcal{P}}\cup \mathcal{I)}.
\end{equation*}%
Then we have the following characterization for anti-zero-sum games. For the
strategy profile $\vec{i}=(i_{p_{1}},i_{p_{2}},\cdots ,i_{p_{n}})$ such that 
$i_{p}\geq 2$ for all $p,$we define 
\begin{eqnarray*}
(E_{\kappa }^{\vec{i}})_{(\hat{\imath}_{p_{1}},\hat{\imath}_{p_{2}},1,\cdots
1)} &=&E_{\kappa }^{(i_{1},i_{2})},(E_{\kappa }^{\vec{i}})_{(\hat{\imath}%
_{p_{1}},\hat{\imath}_{p_{2}},i_{p_{3}},\cdots 1)}=-E_{\kappa
}^{(i_{1},i_{2})},\cdots , \\
(E_{\kappa }^{\vec{i}})_{(\hat{\imath}_{p_{1}},\hat{\imath}%
_{p_{2}},i_{p_{3}},\cdots ,i_{p_{n}})} &=&(-1)^{2n-1}E_{\kappa
}^{(i_{1},i_{2})}
\end{eqnarray*}%
and all other entries are zeros. An example of such tensors for 4 player 2
strategy is given by%
\begin{equation*}
E_{\kappa }^{(2,2,2,2)}=%
\begin{tabular}{l||l}
$%
\begin{pmatrix}
-1 & 1 \\ 
1 & -1%
\end{pmatrix}%
$ & $%
\begin{pmatrix}
1 & -1 \\ 
-1 & 1%
\end{pmatrix}%
$ \\ \hline\hline
$%
\begin{pmatrix}
1 & -1 \\ 
-1 & 1%
\end{pmatrix}%
$ & $%
\begin{pmatrix}
-1 & 1 \\ 
1 & -1%
\end{pmatrix}%
$%
\end{tabular}%
.
\end{equation*}%
Then instead of Proposition \ref{prop:dim-anti-zero}, we will prove the
following proposition.

\begin{proposition}
$\{(E_{\kappa }^{\vec{i}},\cdots E_{\kappa }^{\vec{i}})\}_{\vec{i}\in 
\mathcal{S},\text{ }i_{p}\geq 2\text{ for all }p}$ form a basis for $%
N^{\perp }.$ Thus $\dim (\mathcal{N}^{\perp })=(l-1)^{p}$
\end{proposition}

\begin{proof}
First since $(E_{\kappa }^{\vec{i}},\cdots E_{\kappa }^{\vec{i}})$ is a
symmetric tensor, $\left\langle \,(E_{\kappa }^{\vec{i}},\cdots E_{\kappa }^{%
\vec{i}}),N\right\rangle _{\mathcal{L}^{n}}=0$ for every exact zero-sum game 
$N.$ Also%
\begin{equation*}
\left\langle (E_{\kappa }^{\vec{i}},\cdots E_{\kappa }^{\vec{i}}),(O,\cdots
,E_{\gamma }^{\vec{i}_{-q}},\cdots O)\right\rangle _{\mathcal{L}%
^{n}}=\left\langle E_{\kappa }^{\vec{i}},E_{\gamma }^{\vec{i}%
_{-q}}\right\rangle _{\mathcal{L}}=0.
\end{equation*}%
Thus $\mathrm{span}(\{(E_{\kappa }^{\vec{i}},\cdots E_{\kappa }^{\vec{i}%
})\}_{\vec{i}\in \mathcal{S},\text{ }i_{p}\geq 2\text{ for all }p})\subset $ 
$\mathcal{N}^{\perp }.$ Now we show $\mathcal{N}^{\perp }\subset $ $\mathrm{%
span}$ $(\{(E_{\kappa }^{\vec{i}},\cdots E_{\kappa }^{\vec{i}})\}_{\vec{i}%
\in \mathcal{S},\text{ }i_{p}\geq 2\text{ for all }p}).$ If $%
(A_{p_{1}},\cdots ,A_{p_{n}})\in \mathcal{N}^{\perp },$ then since all $%
(O,\cdots ,Z,\cdots ,-Z,\cdots ,O)\in \mathcal{N},$ $(A_{p_{1}},\cdots
,A_{p_{n}})=(V,\cdots ,V).$ We now show how to express\ $(V,\cdots ,V)$ in
terms of $\{(E_{\kappa }^{\vec{i}},\cdots ,E_{\kappa }^{\vec{i}})\}_{\vec{i}%
\in \mathcal{S},\text{ }i_{p}\geq 2\text{ for all }p}.$ To do this we use an
induction. We suppose that $\{(E_{\kappa }^{\vec{i}},\cdots E_{\kappa }^{%
\vec{i}})\}_{\vec{i}\in \mathcal{S},\text{ }i_{p}\geq 2\text{ for all }p}$
form a basis for the subspace of anti-zero-sum games for $n-1$ player games.
Then for each $i_{p_{n}}\in S$ such that $i_{p_{n}}\geq 2$ (the strategy of $%
n$ th player$),\{(V)_{(i_{p_{1}},i_{p_{2}},\cdots
,i_{p_{n}})}\}_{i_{p_{1}},\cdots ,i_{p_{n-1}}\in S}$ can be viewed as $%
l^{n-1}$ dimensional tensor and hence can be decomposed in terms of a basis
of \ $\{(E_{\kappa }^{\vec{i}},\cdots E_{\kappa }^{\vec{i}})\}_{\vec{i}\in 
\mathcal{S},\text{ }i_{p}\geq 2\text{ for all }p}$ of $n-1$ player games by
the induction hypothesis. In this way we obtain $(l-1)^{n-1}$ coefficients
of the basis elements for each $i_{p_{n}}\geq 2$ and, thus, in total $(\NEG%
{l}-1)^{n}$ coefficients. We write this linear combination as follows:%
\begin{equation*}
B=\sum_{\vec{i}}\kappa ^{\vec{i}}E_{\kappa }^{\vec{i}}
\end{equation*}%
Then we have 
\begin{equation*}
(V)_{(i_{p_{1}},i_{p_{2}},\cdots
,i_{p_{n}})}=(B)_{(i_{p_{1}},i_{p_{2}},\cdots ,i_{p_{n}})}\text{ \ for }%
i_{p_{n}}\geq 2
\end{equation*}%
by construction. Then it follows that $(V)_{(i_{p_{1}},i_{p_{2}},\cdots
,1)}=(B)_{(i_{p_{1}},i_{p_{2}},\cdots ,1)}$ since 
\begin{equation*}
(V)_{(i_{p_{1},}i_{p_{2},\cdots ,}1)}=-\sum_{j\geq
2}(V)_{(i_{p_{1},}i_{p_{2},\cdots ,}j)}=-\sum_{j\geq
2}(B)_{(i_{p_{1},}i_{p_{2},\cdots ,}j)}=(B)_{(i_{p_{1},}i_{p_{2},\cdots
,}1)}.
\end{equation*}
\end{proof}

We illustrate the above proof by the following example. Suppose that $p=2$
and $l=3.$ Suppose that a symmetric bi-matrix game $(A,A)$ is given; $%
A=[a_{1}:a_{2}:a_{3}]$. Then we know that the basis for $\mathcal{N}^{\perp
} $ is given by%
\begin{equation*}
\begin{pmatrix}
-1 & 1 & 0 \\ 
1 & -1 & 0 \\ 
0 & 0 & 0%
\end{pmatrix}%
,%
\begin{pmatrix}
-1 & 1 & 0 \\ 
0 & 0 & 0 \\ 
1 & -1 & 0%
\end{pmatrix}%
,%
\begin{pmatrix}
-1 & 0 & 1 \\ 
1 & 0 & -1 \\ 
0 & 0 & 0%
\end{pmatrix}%
,%
\begin{pmatrix}
-1 & 0 & 1 \\ 
0 & 0 & 0 \\ 
1 & 0 & -1%
\end{pmatrix}%
.
\end{equation*}%
If $A\in \mathcal{N}^{\perp }$ , $\ A$ can be uniquely written as a linear
combination of the above basis. On the other hand, if $A\in \mathcal{N}%
^{\perp },$ then \ $a_{2},$ $a_{3}\in T\Delta $, so $\ a_{2},$ $a_{3}$ can
be uniquely written as a linear combination of $(1,-1,0)^{T},(1,0,-1)^{T}.$
Clearly, the four coefficients that we obtain in the second way also are the
same as the coefficients of the basis elements of $\mathcal{N}^{\perp }.$

\subsection{Proof of Proposition \protect\ref{prop:sym_null_stable}}

\begin{proof}
"If part" is obvious, so we let $A\in \mathcal{L}$ such that $\left\langle
x,PAPx\right\rangle =0$ for all $x\in \mathbb{R}^{l}.$ From the
decomposition we can write $A$ as the following:%
\begin{equation*}
A=\sum_{j\geq i\geq 2}\kappa ^{(ij)}(E_{\kappa }^{(ij)}+E_{\kappa
}^{(ji)})+N+C,\text{ }N\in (\mathcal{M}_{\mathcal{L}})^{\perp },C\in \text{%
ker(}\Gamma )
\end{equation*}%
Since $\left\langle x,PAPx\right\rangle =0$ for all $x\in \mathbb{R}^{l},$
we have%
\begin{equation*}
\sum_{j\geq i}\kappa ^{(ij)}\left\langle x,(E_{\kappa }^{(ij)}+E_{\kappa
}^{(ji)})x\right\rangle =0\text{ for all }x\in \mathbb{R}^{l}.
\end{equation*}%
Let $K^{(ij)}:=\frac{1}{2}(E_{\kappa }^{(ij)}+E_{\kappa }^{(ji)}).$ Next by
choosing appropriate $x$, we show that $\kappa ^{(ij)}=0$ for all $j\geq i.$
Then it follows that $A\in \mathcal{N}_{\mathcal{L}}.$ To do this, observe
that%
\begin{equation*}
\left\langle x,K^{(ij)}x\right\rangle
=-x_{1}^{2}+x_{1}x_{i}+x_{1}x_{j}-x_{i}x_{j},
\end{equation*}%
so whenever $x_{1}=x_{i}$ or $x_{1}=x_{j},\left\langle
x,K^{(ij)}x\right\rangle =0.$We first show that $\kappa ^{(ii)}=0$ for all $%
i.$ For a given $\kappa ^{(mm)},$ we choose%
\begin{equation*}
x=(1,1,\cdots ,1,\overset{m\text{th }}{0},1,\cdots ,1)^{T}
\end{equation*}%
i.e., $x$ is a vector that has $0$ in $n$th element and 1 otherwise. Then
all $i<m,$ $x_{1}=x_{i}=1,$ so $\left\langle x,K^{(ij)}x\right\rangle =0$.
Similarly for all $i>m,$~$x_{1}=x_{i}=1,$ so $\left\langle
x,K^{(ij)}x\right\rangle =0.$ When $i=m,$ since $j>i$, $x_{j}=x_{1}=1,$ thus 
$\left\langle x,K^{(ij)}x\right\rangle =0.$ For $i=j=m,$ $x_{i}=x_{j}=0$ so $%
\left\langle x,K^{(mm)}x\right\rangle =-1.$ Therefore we have $-\kappa
^{(mm)}=0,$which implies $\kappa ^{(mm)}=0.$ Next we show that $\kappa
^{(ij)}=0$ for all $i<j\leq l$ using induction. We start from the highest
index, i.e., $\kappa ^{(l-1,l)}.$ For this case we set 
\begin{equation*}
x=(1,\cdots ,1,\overset{l-1\text{ th }}{0},0)^{T}.
\end{equation*}%
where we assign an arbitrary value to $x_{n}.$ For all $i<l-1,$ $%
x_{1}=x_{i}=1,$ $\left\langle x,K^{(ij)}x\right\rangle =0$ and $\left\langle
x,K^{(l-1,l)}x\right\rangle =-1,$ so $\kappa ^{(l-1,l)}=0.$ Next, we suppose
that $\kappa ^{(ij)}=0$ for all $i>m~\ $and $j>n$ and show that $\kappa
^{(mn)}=0.$ In this case, we set%
\begin{equation*}
x=(1,\cdots ,1,\overset{m\text{ th}}{0},1\cdots ,1,\overset{n\text{ th }}{0}%
,x_{n+1},\cdots ,x_{l})^{T}.
\end{equation*}%
where we assign arbitrary values to elements over $n$th position. Since $n<l$%
, $x\in \mathbb{R}^{l}$. \ For all $i<m,$ $x_{i}=x_{1},\left\langle
x,K^{(ij)}x\right\rangle =0.$ When $i=m$ and $j<n,$ $x_{j}=1,$ so again $%
\left\langle x,K^{(ij)}x\right\rangle =0.$ When $i=m,$ $j=n,$ $%
x_{i}=x_{j}=0. $ Thus $\left\langle x,K^{(ij)}x\right\rangle =-1$ and we
conclude $\kappa ^{(ij)}=0.$
\end{proof}

\subsection{Proof of Proposition \protect\ref{prop:asym_null_stable}}

\begin{proof}
Again "If part" is obvious, so we let $(A,B)\in \mathcal{L}^{2}$ such that $%
\left\langle w,\mathbb{P}(A,B)\mathbb{P}w\right\rangle =0$ for all $w\in 
\mathbb{R}_{l_{r}+l_{c}}.$ From corollary \ref{cor:decomp} (3), \ we can
write $(A,B)$ as 
\begin{equation*}
(A,B)=\sum_{i\geq 2,j\geq 2}\kappa ^{(ij)}(E_{\kappa }^{(ij)},E_{\kappa
}^{(ij)})+(N,-N)+(C_{1},C_{2}),\text{ }
\end{equation*}%
where $(N,-N)\in \mathcal{M}^{\perp },(C_{1},C_{2})\in \ker (\mathbf{\Gamma }%
).$ Since $\left\langle w,\mathbb{P}(A,B)\mathbb{P}w\right\rangle =0$ for
all $w\in \mathbb{R}_{l_{r}+l_{c}},$ we have%
\begin{equation*}
\sum_{i\geq 2,j\geq 2}\kappa ^{(ij)}(\left\langle y,(E_{\kappa
}^{(ij)})^{T}x\right\rangle _{\mathcal{L}}+\left\langle x,E_{\kappa
}^{(ij)}y\right\rangle _{\mathcal{L}})=0\text{ for all }x\in \mathbb{R}%
^{l_{r}},\text{ }y\in \mathbb{R}^{l_{c}}.
\end{equation*}%
Similarly to the previous section, by choosing appropriate $x$ and $y$ we
show that $\kappa ^{(ij)}=0$ for all $i\geq 2,j\geq 2.$ Then it follows that 
$(A,B)\in \mathcal{N}.$ To do this, observe that%
\begin{equation}
\frac{1}{2}(\left\langle y,E_{\kappa }^{(ji)}x\right\rangle _{\mathcal{L}%
}+\left\langle x,E_{\kappa }^{(ij)}y\right\rangle _{\mathcal{L}%
})=-x_{1}y_{1}+x_{i}y_{1}+x_{1}y_{j}-x_{i}y_{j},  \label{eq:null-st}
\end{equation}%
so whenever $x_{1}=x_{i}$ or $y_{1}=y_{j},$(\ref{eq:null-st}) becomes zero$.$
We choose the following $(x^{(i)},y^{(j)})$:%
\begin{equation*}
x^{(i)}=(1,1,\cdots ,1,\overset{i\text{th }}{0},1,\cdots ,1)^{T},\text{ }%
y^{(j)}=(1,1,\cdots ,1,\overset{j\text{th }}{0},1,\cdots ,1)^{T}.
\end{equation*}%
Then for $(k,m)$ such that $k\neq i$ or $m\neq j,$ we have either $%
x_{1}^{(i)}=x_{k}^{(i)}$ or $y_{j}^{(j)}=y_{1}^{(j)}.$ Thus for all $(k,m)$
such that $k\neq i$ or $m\neq j,$%
\begin{equation*}
\left\langle y^{(j)},E_{\kappa }^{(mk)}x^{(i)}\right\rangle _{\mathcal{L}%
}+\left\langle x^{(i)},E_{\kappa }^{(km)}y^{(j)}\right\rangle _{\mathcal{L}%
}=0
\end{equation*}%
and%
\begin{equation*}
\left\langle y^{(j)},E_{\kappa }^{(ji)}x^{(i)}\right\rangle _{\mathcal{L}%
}+\left\langle x^{(i)},E_{\kappa }^{(ij)}y^{(j)}\right\rangle _{\mathcal{L}%
}=-2.
\end{equation*}%
From this we conclude that $\kappa ^{(ij)}=0.$ Thus, $(A,B)\in \mathcal{N}.$
\end{proof}

\subsection{Proof of Corollary \protect\ref{cor:3_st_strict_stable}}

\begin{proof}
From proposition \ref{prop:strict-stable}, we see that $[A]$ is strictly
stable if and only if $S,$ $A$'s part belonging to $\mathcal{N}_{\mathcal{L}%
}^{\perp },$ is strictly stable and $S$ has the following parameterization. 
\begin{equation*}
S=%
\begin{pmatrix}
-a-b & a & b \\ 
a & -a-c & c \\ 
b & c & -b-c%
\end{pmatrix}%
\end{equation*}%
We recall that $\left\langle x,Sx\right\rangle $ satisfying $\sum_{i}x_{i}=0$
is negative if and only if its bordered Hessians, given below, satisfies
some sign condition as we will check below. In our case, these conditions are%
\begin{equation*}
\det 
\begin{pmatrix}
-a-b & a & 1 \\ 
a & -a-c & 1 \\ 
1 & 1 & 0%
\end{pmatrix}%
>0,\text{ \ }\det 
\begin{pmatrix}
-a-b & a & b & 1 \\ 
a & -a-c & c & 1 \\ 
b & c & -b-c & 1 \\ 
1 & 1 & 1 & 0%
\end{pmatrix}%
\text{\ }<0.
\end{equation*}%
Then by computing determinants we find that 
\begin{equation*}
4a+b+c>0\text{ and }ab+bc+ca>0
\end{equation*}%
and obtain the desired result.
\end{proof}

\subsection{Proof of Proposition $\protect\ref{prop:con-vol}$}

\begin{proof}
(1) First we note that $x^{T}Ax=0~$\ and $A\mathbf{1}=0\mathbf{,}$ so $%
x_{0}=(\frac{1}{n},\cdots ,\frac{1}{n})$ is a rest point for ($\ref{rep}).$
We consider $H(x):=\sum_{i}\log (x_{i}).$ Then $LH=\sum (Ax)_{i}=0,$ thus $H$
is an integral of (\ref{rep}). Thus (\ref{rep}) is conservative. To show the
preservation of volume we first write $\hat{x}=(1-\sum_{i\neq
1}x_{i},x_{2},\cdots ,x_{n})$ and when $x\in \Delta ,Ax=A\hat{x}$ and $%
\left\langle x,Ax\right\rangle =\left\langle \hat{x},A\hat{x}\right\rangle .$
Also we note that for $k\geq 2,$%
\begin{eqnarray*}
\frac{\partial }{\partial x_{k}}(A\hat{x})_{k} &=&-a_{k1}+a_{kk} \\
\frac{\partial }{\partial x_{k}}\left\langle \hat{x},A\hat{x}\right\rangle
&=&-(Ax)_{1}-(A^{T}x)_{1}+(A^{T}x)_{k}+(Ax)_{k}
\end{eqnarray*}%
Thus 
\begin{eqnarray*}
\func{div}_{\Delta }f_{A} &=&\sum_{k\neq 1}\frac{\partial f_{k}}{\partial
x_{k}}(x)=\sum_{k\neq 1}(Ax)_{k}-(l-1)\left\langle x,Ax\right\rangle
-\sum_{k\neq 1}x_{k}a_{k1}+\sum_{k\neq 1}x_{k}a_{kk} \\
&&+(1-x_{1})(Ax)_{1}+(1-x_{1})(A^{T}x)_{1}-\sum_{k\neq
1}x_{k}(A^{T}x)_{k}-\sum_{k\neq 1}x_{k}(Ax)_{k} \\
&=&\sum_{k}(Ax)_{k}-l\left\langle x,Ax\right\rangle
+\sum_{k}x_{k}a_{kk}-\left\langle x,A^{T}x\right\rangle
\end{eqnarray*}%
If $A$ is anti-potential, then $\sum_{k}(Ax)_{k}=\left\langle \mathbf{1}%
,Ax\right\rangle =\left\langle A^{T}\mathbf{1},x\right\rangle =0$ and all
diagonal elements of $A$ are zero. Thus $\func{div}_{\Delta }f_{A}=0$\newline
(2)\ Recall that $(A,B)$ $\in range(\mathbf{\Gamma })$ if and only if $(%
\mathbf{1}_{r}^{T}A,\mathbf{1}_{r}^{T}B)=0$ and $(A\mathbf{1}_{c},B\mathbf{1}%
_{c})=0.$Thus $\left( A,B\right) $ has an interior rest point, so from %
\citet{Hofbauer98} (p.130) the result follows.\newline
(3) Similarly we have, for $l_{r}=:l\geq 2$%
\begin{equation*}
\frac{\partial }{\partial x_{l}}\left\langle \hat{x},Ay\right\rangle
=(Ay)_{l}-(Ay)_{1},\text{ and}~\ \frac{\partial }{\partial y_{l}}%
\left\langle \hat{y},B^{T}x\right\rangle =(B^{T}x)_{l}-(B^{T}x)_{1}.
\end{equation*}%
Thus%
\begin{eqnarray*}
\func{div}_{\Delta }f_{(A,B)} &=&\sum_{i\neq 1}\frac{\partial f_{i}}{%
\partial x_{i}}(x,y)+\sum_{j\neq 1}\frac{\partial f_{j}}{\partial y_{j}}%
(x,y)=\sum_{i\neq 1}((Ay)_{i}-\left\langle x,Ay\right\rangle )-\sum_{i\neq
1}x_{i}((Ay)_{i}-(Ay)_{1}) \\
&&+\sum_{j\neq 1}((B^{T}x)_{i}-\left\langle y,B^{T}x\right\rangle
)-\sum_{j\neq 1}y_{j}((B^{T}x)_{j}-(B^{T}x)_{1}) \\
&=&\sum_{i}(Ay)_{i}-l_{r}\left\langle x,Ay\right\rangle
+\sum_{j}(B^{T}x)_{j}-l_{c}\left\langle y,B^{T}x\right\rangle .
\end{eqnarray*}%
Then since $\left( A,B\right) $ is anti-symmetric in $\mathcal{L}^{2},$
which implies $\left\langle x,Ay\right\rangle +\left\langle
y,B^{T}x\right\rangle =0,$and $\left( A,B\right) \in range(\mathbf{\Gamma ),}
$ the result follows.
\end{proof}

%\newpage

\bibliographystyle{elsarticle-harv}
\bibliography{evolutionary_games}

\end{document}